\documentclass{article}

\usepackage{PRIMEarxiv}

\usepackage[utf8]{inputenc} 
\usepackage[T1]{fontenc}    
\usepackage{hyperref}       
\usepackage{url}  
\usepackage[inkscapelatex=false]{svg}
\usepackage{booktabs}       
\usepackage{amsfonts}       
\usepackage{nicefrac}       
\usepackage{microtype}      
\usepackage{lipsum}
\usepackage{amsmath}
\usepackage{authblk}
\usepackage{amssymb}
\usepackage{algorithm}
\usepackage{algpseudocode}
\usepackage{caption} 
\usepackage{fancyhdr}       
\usepackage{graphicx}       
\graphicspath{{media/}}     
\DeclareMathOperator*{\argmax}{arg\,max}
\title{PyGen: A Collaborative Human-AI Approach to Python Package Creation}

\author[1,a]{Saikat Barua}
\author[2]{Mostafizur Rahman}
\author[3]{Md Jafor Sadek}
\author[4]{Rafiul Islam}
\author[5]{Shehenaz Khaled}
\author[6]{Dr. Md. Shohrab Hossain}

\affil[1]{North South University, Dhaka, \texttt{saikat.barua@northsouth.edu}}
\affil[a]{SpontAlign, Dhaka}
\affil[2]{North South University, Dhaka, \texttt{mostafizur.rahman10@northsouth.edu}}
\affil[3]{North South University, Dhaka, \texttt{jaforsadek619@northsouth.edu}}
\affil[4]{North South University, Dhaka, \texttt{rafiul.islam19@northsouth.edu}}
\affil[5]{North South University, Dhaka, \texttt{shehenaz.khaled@northsouth.edu}}
\affil[6]{Bangladesh University of Engineering and Technology, Dhaka, \texttt{mshohrabhossain@cse.buet.ac.bd}}

\begin{document}
\maketitle

\begin{abstract}

The principles of automation and innovation serve as foundational elements for advancement in contemporary science and technology. Here, we introduce Pygen, an automation platform designed to empower researchers, technologists, and hobbyists to bring abstract ideas to life as core, usable software tools written in Python. Pygen leverages the immense power of autoregressive large language models to augment human creativity during the ideation, iteration, and innovation process. By combining state-of-the-art language models with open-source code generation technologies, Pygen has significantly reduced the manual overhead of tool development, thereby significantly enhancing creativity and productivity. From a user prompt, Pygen automatically generates Python packages for a complete workflow from concept to package generation and documentation. Practical examples of libraries such as AutoML, AutoVision, AutoSpeech, and Quantum Error Correction are demonstrated. The findings of our work show that Pygen considerably enhances the researcher's productivity by enabling the creation of resilient, modular, and well-documented packages for various specialized purposes. We employ a prompt enhancement approach to distill the user's package description into increasingly specific and actionable. While being inherently an open-ended task, we have evaluated the generated packages and the documentation using Human Evaluation, LLM-based evaluation, and CodeBLEU, with detailed results in the results section. Furthermore, we documented our results, analyzed the limitations, and suggested strategies to alleviate them. Pygen is our vision of ethical automation, a framework that promotes inclusivity, accessibility, and collaborative development. This project marks the beginning of a large-scale effort towards creating tools where intelligent agents collaborate with humans to improve scientific and technological development substantially.

Our code and generated examples are open-sourced at \url{https://github.com/GitsSaikat/Pygen}

\end{abstract}


\section{Introduction}

Curiosity, innovation, and relentless pursuit have always characterized scientific progress. Today, we stand on the threshold of a promising new chapter in which every step, though minor and significant, acts as a booster for scientific growth. Pygen is a system that represents a significant stride toward that vision. It aims to automate humdrum and recurrent activities to free time for researchers, scientists, and enthusiasts to practice what matters: creativity and breakthrough innovation effectively. Automation becomes one of the most vital tools for social benefit when used thoughtfully and diligently\cite{carayannis2012}. Automation simplifies tasks, opens vistas for reimagining processes, and makes those processes even better across disciplines. Treated right and with responsibility, it might be the beacon of progress for everyone\cite{brynjolfsson2014}. Pygen do just that: make technology accessible, amply productive, and take people to new heights in their scientific journeys as technology empowers individuals to achieve their milestones along with the progress of civilization\cite{boden2004}.

The path to innovation involves solving unique and complex problems. Not all challenges can be addressed with existing technology, but human creativity shines through in its ability to construct new tools as needed, thereby expanding the boundaries of what is possible\cite{hwang2022too}\cite{shneiderman2000creating}. Pygen embodies this spirit of innovation by transforming abstract ideas into practical solutions that make a tangible impact. When the scientific community tackles a problem and finds a solution, they often discover key components and experimental techniques that are valuable for future scientists and technologists\cite{barrett2020}. Creating Python packages originated from the desire to equip the community with essential tools that streamline experimental processes and advance scientific work. This approach to building tools is a form of responsible automation, where the human element remains integral, allowing for flexibility, creativity, and adaptability—qualities that purely automated tool designs often lack.

Our vision for Pygen is to create a dynamic system that helps generate effective software tools and nurtures and inspires new ideas. By focusing on responsible automation, we aim to empower researchers and technologists to explore new possibilities, build upon existing knowledge, and contribute to advancing science and technology. Traditional software automation\cite{Alvarez2018A}\cite{Jackson2006Software}\cite{Shaw1990Toward} approaches primarily focus on creating user-level abstractions that integrate a user's perspective to improve the software. However, our approach emphasizes the importance of designing superior tools that lead to the creation of better-end products. We made a critical observation that humans, when faced with challenges, often develop new tools if existing ones are inadequate\cite{stout2007neuroscience}\cite{nielsen2014exploring}. Likewise, a language-model-based agentic system tasked with complex work must not only learn to use available tools but also create new ones to solve problems effectively\cite{Cai2023Large}\cite{Qin2023Tool}\cite{Paisner2014Goal-Driven}. This insight led us to develop an automated Python package generation system that starts with simple user prompts and evolves from there. By integrating this approach into an agentic framework, Pygen aims to enhance the adaptability and performance of such systems, enabling them to tackle increasingly sophisticated tasks and deliver impactful results.

Foundational models\cite{anthropic2024}\cite{googledeepmind2023}\cite{llama2024}\cite{openai2023} have typically been used to generate code for direct use, but their potential to autonomously build tools and design comprehensive software solutions remains largely untapped. While these models can assist in writing scripts or automating simple tasks, they have yet to augment human capability in complex, multi-faceted project settings effectively. With Pygen, we aim to change this paradigm. We are empowering these foundational models to generate ideas for software tools and create Python packages that can be effectively used to solve real-world challenges. By producing thorough documentation alongside the generated tools, Pygen extends the capabilities of these models beyond simple automation, turning them into meaningful partners in creative and technical endeavors.

The concept of the AI scientist\cite{lu2024aiscientist} inspires our work, which is an end-to-end framework capable of originating novel ideas, developing experiments to explore those ideas, and ultimately producing scientific literature to share the resulting insights. Pygen builds on this vision by enhancing user queries, generating Python packages, and providing comprehensive documentation that allows others to easily understand, utilize, and build upon the generated tools. This process empowers users by transforming abstract ideas into functional, well-documented tools, simplifying the journey from initial concept to practical application. Pygens do more than merely automate; they act as extensions of human creativity. It bridges the gap between high-level conceptual thinking and practical, hands-on implementation, allowing users to bring their ideas to life with minimal friction.

Pygen allows users to specify the type of package they need for their tasks, along with the desired features and functionalities. Based on the user's description, the system refines these ideas and creates optimized implementation strategies. Using these refined strategies, Pygen designs a Python package by leveraging open-source models available on platforms like Google AI Studio and the GroqCloud. Once the package is generated, the Pygen creates comprehensive documentation to accompany it. Users can download the package and its documentation as a zip file, ensuring that all necessary information is in one place. The package is automatically set up if executed in the user's local environment, allowing for a smooth transition from development to execution. Users can further enhance these packages to meet their specific needs and, if desired, deploy them within the Python ecosystem.

A key principle behind Pygen is our emphasis on open-source accessibility. Using open-source models, we ensure that users can access the system without being hindered by financial barriers or paywalls. This approach promotes open access and open-source scientific discovery, allowing individuals from all backgrounds to contribute to and benefit from innovation. Thanks to this open-source pipeline, users can utilize models made available through platforms such as GroqCloud and Google AI Studio to generate and document Python packages completely free of charge, encouraging experimentation and continuous improvement of Pygen. We are committed to open-sourcing Pygen itself, inviting contributions from the broader community to enhance its capabilities further, making it a truly collaborative and evolving project.

Our contributions are summarized as follows: 

\begin{enumerate}
    \item  We have introduced a Python package development system powered by open-source frontier models. This pipeline transforms user descriptions into refined ideas, leading to the generation of Python packages accompanied by thorough documentation. Users can download the generated package and documentation seamlessly, enabling them to start working immediately.

    \item Pygen can be deployed as a user-friendly application, allowing users to access it directly by simply setting their API key. This streamlined access reduces barriers to entry and enables a broader audience to leverage the system's capabilities.

    \item We have outlined several future research directions to improve responsible system automation. These include refining the strategies for improving Pygen, integrating a package reviewer to ensure robustness and reliability, and exploring the potential of agentic frameworks that can autonomously create and refine the tools they use.
    
\end{enumerate}

\begin{figure}[htbp]
    \centering
    \includegraphics[width=0.45\linewidth]{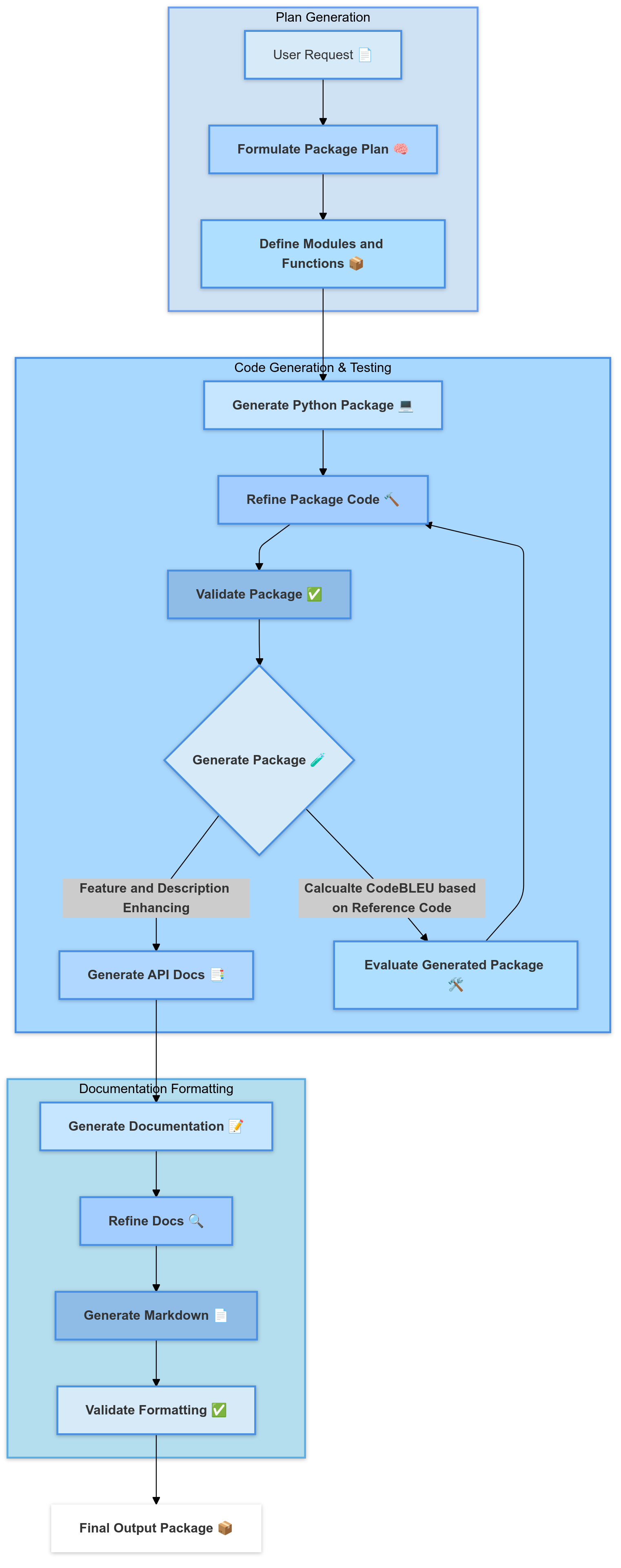} 
    \caption{Pygen's Workflow: This diagram describes Pygen's workflow to generate a Python package given the user's request. First, it starts with generating the plan, which includes formulating some package plan according to the user's needs. It then involves users' prompt refinement, based on validating, generating, and evaluating a package. The final step is documentation generation and formatting. At the end, documents are created, refined, and converted to Markdown to enable formatting validation for the final output package.}
    \label{fig:Pygen_workflow}
\end{figure}

\section{Background} 

\paragraph{Large language models} In this paper, we present an automated scientist system that leverages the advanced capabilities of autoregressive large language models. Specifically, we utilize sophisticated models like those developed by Google DeepMind's Gemini Team\cite{team2023gemini} and the Llama Team \cite{touvron2023llamaopenefficientfoundation}. These models are designed to generate predictive text completions by estimating the probability distribution of the next token—akin to a word—based on the sequence of preceding tokens. This process is mathematically represented as $p(x_t \mid x_{<t}; \theta)$, where the model calculates the conditional probability of each potential next token. A sampling procedure follows this estimation phase during the testing phase, which generates coherent and contextually appropriate outputs. 

The underlying power of these large language models (LLMs) stems from their training on vast and diverse datasets, which equips them to generate fluent and contextually relevant text and exhibit a wide range of advanced, human-like abilities. For instance, these models can effectively leverage commonsense knowledge\cite{touvron2023llamaopenefficientfoundation}, perform logical and abstract reasoning tasks\cite{wei2022chain}, and even write, interpret, and debug complex code structures. The ability to generate such a diverse set of outputs highlights the adaptability and sophistication of these models when they are properly scaled and trained with extensive datasets. This adaptability makes LLMs a versatile tool for tackling various tasks, from natural language processing to aiding in scientific research.

\paragraph{GroqCloud} In practical applications, we have used GroqCloud, a platform that enables users to utilize different open-source models, including Meta's Llama and Google's Gemini, among others, such as the Mistral model\cite{mistral2023}. Their API can give much faster inference times than manually achievable, making deploying more complex language models less onerous and technically demanding. GroqCloud is designed to be an easy entry point for researchers and developers looking to harness the power of strong AI models and simplify everything from the beginning into actual use. GroqCloud democratizes access to sophisticated AI thanks to the abstractions brought in by removing heavy computational needs and exhaustive model tuning.

\paragraph{Local Hosting} Besides showing cloud-based solutions, we have also shown how to use Ollama for downloading and self-serving the model. This provides full integrations for users with tools like Pygen for key components to facilitate easy execution while maintaining full customization locally and privacy. Hosting local models using Ollama will enable developers to solve probable problems usually associated with security and latency in cloud-based systems.

\section{Related Works}

Integrating Large Language Models (LLMs) in complex domains has significantly advanced their use in various applications, such as scientific discovery, multi-agent collaboration, and automated reasoning. Lu et al. (2024) introduce The AI Scientist, a fully automated framework for scientific discovery, capable of independently generating hypotheses, conducting experiments, and producing research papers that meet top conference standards, marking a significant leap in automating the scientific process\cite{lu2024aiscientist}. Yang et al. (2022) presented LLMs as inductive reasoners, addressing the limitations of formal languages in reasoning tasks \cite{Yang2022LanguageModels}. Zhong et al. (2023) introduced a goal-driven discovery system, which identified distributional differences using language descriptions, showcasing the potential of LLMs for efficient data-driven research \cite{Zhong2023GoalDrivenDiscovery}. Zhou et al. (2024) demonstrated the evolution of self-evolving agents through symbolic learning, emphasizing the transition from model-centric to data-centric systems \cite{Zhou2024SymbolicLearning}. Chen et al. (2023) explored multi-agent frameworks and emergent behaviors, emphasizing how collaborative models can dynamically adapt \cite{Chen2023AgentVerse}. Song et al. (2024) proposed a new adaptive team-building paradigm for multi-agent systems to solve complex tasks flexibly \cite{Song2024AdaptiveTeam}. In the survey paper on autonomous agents \cite{barua2024exploring}, the operational mechanisms of LLM-based agents are thoroughly explored and explained.

Gu and Krenn (2024) highlighted using knowledge graphs and LLMs for generating compelling interdisciplinary research ideas, assessed by 100 research group leaders \cite{Gu2024ScientificIdeaGen}. Xu et al. (2023) proposed MAgIC, which benchmarks LLM agents on adaptability and collaboration, revealing substantial variations across different models \cite{Xu2023MAgIC}. Qin et al. (2023) introduced ToolLLM, a framework enabling LLMs to utilize over 16,000 real-world APIs effectively, significantly bridging the gap between LLMs and real-world applications \cite{Qin2023ToolLLM}. Yuan et al. (2024) presented EvoAgent, an evolutionary algorithm to automatically extend expert agents into multi-agent systems, enhancing problem-solving capabilities \cite{Yuan2024EvoAgent}. Arora et al. (2024) introduced MASAI, a modular architecture for software engineering agents, allowing sub-agents to solve distinct sub-problems effectively \cite{Arora2024MASAI}. 

Si et al. (2024) analyzed the novelty of research ideas generated by LLMs, revealing that while these ideas are more novel, human experts still surpass LLMs in terms of feasibility \cite{Si2024CanLLMsGenerate}. Chen et al. (2024) proposed AutoManual, a framework for LLMs to autonomously generate instruction manuals by interacting with and learning from environments \cite{Chen2024AutoManual}. Zhuge et al. (2024) described LLM-based agents as optimizable graphs capable of automatic graph optimization to improve multi-agent collaboration \cite{Zhuge2024LanguageAgents}. Gao et al. (2024) presented AgentScope, a multi-agent platform designed to enhance robustness and coordination among agents \cite{Gao2024AgentScope}. Wang et al. (2023) focused on optimizing novelty in scientific inspiration through SciMON, a framework designed to generate research ideas grounded in scientific literature \cite{Wang2023SciMON}. 

Kumar et al. (2024) evaluated LLMs' ability to generate novel research ideas and highlighted the promising role of these models in contributing to interdisciplinary research \cite{Kumar2024UnlockResearch}. Yang et al. (2023) proposed an automated open-domain hypothesis discovery system, emphasizing the capability of LLMs to propose novel, valid hypotheses without pre-existing human annotations \cite{Yang2023HypothesisDiscovery}. Wang et al. (2024) introduced CodeAct, which uses executable Python code as action space for LLM agents, facilitating dynamic interaction with environments \cite{Wang2024ExecutableActions}. Qi et al. (2023) presented LLMs as zero-shot hypothesis proposers, demonstrating that these models can generate valid scientific hypotheses even without prior training \cite{Qi2023ZeroShotHypothesis}. Sprueill et al. (2024) introduced ChemReasoner, which combines linguistic reasoning with quantum-chemical feedback for catalyst discovery, showing a promising pathway for AI-assisted chemistry \cite{Sprueill2024ChemReasoner}. 

\begin{figure}[htbp]
    \centering
    \includegraphics[width=0.8\linewidth]{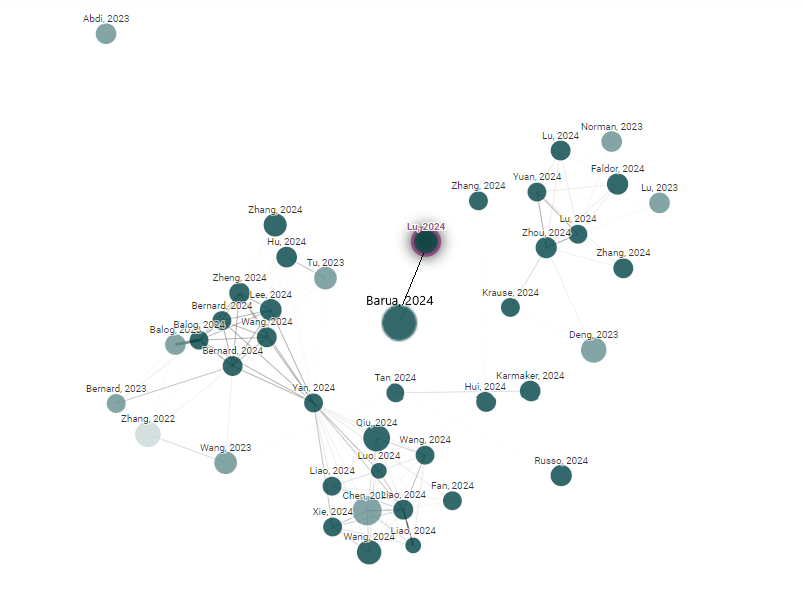} 
    \caption{Literature Network Map: This diagram shows connections among various research publications, illustrating the citation relationships between different authors. Larger nodes represent influential papers, while the links depict citations connecting related works. The clustering of nodes provides insights into research communities and key contributions in the field. The large central node presents our paper.}
    \label{fig:Literature_Network_Map}
\end{figure}

Fernando et al. (2023) proposed Promptbreeder, a self-improvement mechanism for LLM prompts, achieving enhanced performance on reasoning benchmarks \cite{Fernando2023Promptbreeder}. Anderson et al. (2024) analyzed the homogenization effect of LLMs on creative ideation, noting that LLM users generated less distinct ideas than traditional creativity support tools \cite{Anderson2024Homogenization}. Yin et al. (2024) introduced Gödel Agent, a self-referential agent framework designed for recursive self-improvement \cite{Yin2024GodelAgent}. Li et al. (2024) proposed MLR-Copilot, a machine learning research assistant that uses LLMs for automatic research generation and implementation \cite{Li2024MLRCopilot}. Hu et al. (2024) demonstrated the automated design of agentic systems using meta-agent programming, highlighting the development of novel agents through a meta-agent framework \cite{Hu2024AutomatedDesign}. Figure\ref{fig:Literature_Network_Map} shows the overview of the literature review.

\section{The PyGen}

\paragraph{Overview} The Pygen has three main phases: Plan Generation, Code Generation and Testing, and Documentation Generation. In general, Pygen systematically transforms a user's requirement into a full Python package following a structured approach. First, the Phase of Plan Generation is all about understanding and scoping the package and formulating a detailed package plan that fits the user's needs. During this stage, Pygen takes the natural language inputs given by the user to specify the modules and functions that will make up the backbone of the package. Therefore, this is a very important step in providing a sound basis for the rest of the development activities by converting the conceptual requirements into actionable technical specifications.
The generation of code and test cases occurs in the second stage, based on the planning from the first step. Functional code can be generated in Pygen using advanced language models. If already predefined, then modules and functions can be used to create functional code. Refinement might still be necessary after the code has been generated to increase quality and verify that the overall package structure is sound. The refinement steps are provided by enhancing the feature descriptions provided by the user. In the final step, documentation is created to ensure the delivered package is high quality and user-friendly. Pygen generates documentation from the generated package automatically. Subsequently, documents are transformed into Markdown format for easier reading and distribution. The Markdown documents are checked through formatting validation to ensure the standards necessary for user accessibility and consistency. The outcome of this stage is a well-documented package ready to use and distribute. The complete workflow of Pygen is depicted in Figure \ref{fig:Pygen_workflow}.

\subsection{Plan Generation}

Generating a plan is one of the most crucial steps in the Pygen workflow and starts the effective and efficient creation of Python packages. Our work has been greatly inspired by evolutionary computation and open-endedness research\cite{lu2024aiscientist}\cite{brant2017minimal}\cite{stanley2017open}\cite{lehman2022evolutionlargemodels}. Pygen initiates this phase by interacting with the user to get detailed input about the package's needs. The user provides a prompt with a small description of the package and a set of features that shall be included. These will form the basis of the intended functionality of the package. That phase of the generation process is iterative: The users refine and expand their descriptions and lists of features until those become comprehensive and specific. Once Pygen has collected the details for each feature, thanks to its powerful set of language model capabilities-driven generation tools, such as the Groq client, detailed, user-friendly yet technically informative descriptions are generated for each feature. This often includes supporting pseudocode or implementation hints that help in subsequent coding. Pygen also employs retry mechanisms and exponential backoff for robust communication with the language model to handle temporary network or service issues effectively. In this manner, Pygen takes high-level user inputs into actionable technical specifications that seamlessly move into the coding phase. We have implemented a prompt-enhancing approach in a manner that resonates with prompt tuning research\cite{lester2021power}\cite{liu2021p}\cite{zhu2023prompt}\cite{jia2022visual}.

The second aspect of the Plan Generation phase is enhancing the overall package description via prompt caching\cite{zhang2023prompt}\cite{gim2024prompt}. Pygen refines this into an approachable, helpful package for developers with more complete implementation details; it does this through multiple iterations of interaction with the language model and hence is highly informative and ready for downstream tasks. Pygen will ensure feature descriptions and the entire package description are persisted for later summarization - once more by the model - in JSON and text file formats to create a prompt context for the next step in package generation. In summary, Plan Generation generally emphasizes transforming abstract user ideas into concrete and structured plans through natural language processing and iterative refinement.

\subsection{Package Creation}

The package creation process begins by gathering key inputs, including an enhanced package description and feature specifications from the previous step. These elements are then processed by sophisticated language models to generate the package structure and content, adhering to a predefined template. The template is designed based on the specifications of the Python Standard Library\cite{hellmann2011python}.  Leveraging these models ensures a high level of natural language understanding, facilitating the transformation of user-defined requirements into structured, well-documented Python code. The package generation includes essential files and module stubs, ensuring compliance with established Python distribution standards. The generated package is initially saved in the local environment and can be refined or downloaded as a zip file based on user requirements.

Pygen's capabilities are enhanced by integrating a high-performance platform that supports open-source models optimized for computational efficiency, scalability, and faster inference\cite{ahmed2022answer}\cite{abts2022software}. This approach leverages specialized models to manage requests and generate the package structure and content, significantly improving model availability and response times. The system generates comprehensive packages that include necessary scripts, test files, and documentation by interpreting user prompts and converting them into actionable Python package structures. This approach allows users to utilize more complex and powerful models while adhering to a strict package content and structure format. Leveraging specialized tools ensures that the generated packages are optimized for environments where rapid iteration and resource efficiency are crucial, highlighting Pygen's adaptability and commitment to efficient and effective package generation.

Lastly, Pygen offers a self-hosted model generation pathway, enabling local hosting and interaction with models using Ollama after initial setup. This flexibility makes it ideal for scenarios prioritizing data privacy, autonomy, or low latency. Users retain complete control over their data by performing the entire package generation process locally while leveraging sophisticated language model capabilities. This workflow, while similar to other approaches, strongly emphasizes secure, offline usage. The installation and setup instructions make this pathway accessible to developers seeking customized environments for tool creation.

\subsection{Creating Documentation and Formatting}

In Pygen, automated documentation is an essential ingredient; it has been vital in delivering packages that are accessible and usable to Python. The focus here is turning the structured output of a generated package into a documented one. Our documentation-generating template is designed based on insights from best practices\cite{Zhi2015Cost}. This generation process starts by extracting critical information from the descriptive content, setup configurations, dependencies, and usage scenarios from various package components into one coherent documentation framework. This, therefore, becomes a detailed guide that encompasses how to use, install, and contribute to the package with features, examples, and API references, among others. Besides promoting a more excellent user experience, extensive documentation contributes to the broader community by providing deeply detailed, reusable information about the created tools.

For a start, some of the essential elements of this automated documentation will include an overview that contains the package descriptions, which are deduced from general descriptive content; a description of the features of the package deduced from configuration details; sections on installation, usage examples, API references, and testing protocols, step by step demonstration of how to use the library, by using a language model's capability to parse classes, functions, and dependencies, generated documentation results in enrichment with technical specifics but easy readability. Contribution guidelines and licensing information further encourage community involvement and ensure that the generated packages comply with open-source principles. This automated yet detailed documentation strategy makes Pygen packages easy to understand and ready for integration into broader scientific and development workflows that significantly reduce manual overhead, which is usually needed for comprehensive software documentation.

\section{Mathematical Preliminaries}

We present the mathematical foundations underlying Pygen's core components: language model inference, documentation generation, and package structure optimization. These formalisms provide the theoretical framework for understanding how Pygen transforms user requirements into functional Python packages.

\subsection{Language Model Foundations}

The foundation of Pygen relies on autoregressive language models that generate text by estimating conditional probabilities. Given a sequence of tokens $x = (x_1, ..., x_T)$, the model computes the probability of each token given its preceding context:

\begin{equation}
    p(x) = \prod_{t=1}^T p(x_t | x_{<t})
\end{equation}

where $x_{<t}$ represents all tokens before position $t$. The model parameters $\theta$ are learned through training to minimize the negative log-likelihood:

\begin{equation}
    \mathcal{L}(\theta) = -\sum_{t=1}^T \log p(x_t | x_{<t}; \theta)
\end{equation}

Modern LLMs, including those utilized by Pygen, are predominantly based on the Transformer architecture \cite{vaswani2017attention}. The Transformer leverages self-attention mechanisms to model dependencies between tokens irrespective of their positional distances. Mathematically, the self-attention mechanism computes the attention scores between pairs of tokens using the following formulation:

\begin{equation}
    \text{Attention}(Q, K, V) = \text{softmax}\left(\frac{QK^{\top}}{\sqrt{d_k}}\right)V
\end{equation}

where \( Q \), \( K \), and \( V \) denote the query, key, and value matrices, respectively, and \( d_k \) is the dimensionality of the key vectors. This mechanism allows the model to weigh the relevance of different tokens dynamically, facilitating the capture of complex linguistic and contextual relationships. The optimization of LLMs is typically performed using variants of stochastic gradient descent (SGD), such as Adam \cite{kingma2014adam}, which adjusts the learning rates adaptively based on the first and second moments of the gradients. Regularization techniques, including dropout\cite{srivastava2014dropout} and weight decay, are employed to prevent overfitting and enhance the generalization capabilities of the model.

During inference, Pygen uses temperature sampling to control the randomness of generated outputs. For a temperature parameter $\tau > 0$, the sampling probability is computed as:

\begin{equation}
    p_\tau(x_t | x_{<t}) = \frac{\exp(\log p(x_t | x_{<t})/\tau)}{\sum_{x'} \exp(\log p(x' | x_{<t})/\tau)}
\end{equation}

A lower temperature (\( \tau < 1 \)) makes the model more deterministic, favoring higher-probability tokens, while a higher temperature (\( \tau > 1 \)) increases randomness, allowing for more diverse outputs \cite{vaswani2017attention, radford2019language}.

\subsection{Package Structure Organization}

Pygen optimizes package structure using a hierarchical representation. Let $G = (V, E)$ be a directed acyclic graph where:
\begin{itemize}
    \item $V$ represents the set of package components (modules, functions, classes)
    \item $E$ represents dependencies between components
\end{itemize}

The optimal package structure minimizes the objective function:

\begin{equation}
    \min_{G} \sum_{(u,v) \in E} w(u,v) + \lambda \sum_{v \in V} c(v)
\end{equation}

In this context, \( w(u,v) \) represents the coupling weight between components, \( c(v) \) denotes the complexity of component \( v \), and \( \lambda \) is a regularization parameter. In algorithm 1, the package generation workflow is elaborated.

\begin{algorithm}
\caption{Python Package Generation}
\begin{algorithmic}[1]

    \Procedure{get\_api\_key}{}
        \State Return API key from env or prompt.
    \EndProcedure
    
    \Procedure{parse\_content}{content}
        \State Initialize files dict, state to `expect\_file`.
        \For{line in content}
            \If{expecting file path \& valid}
                \State Set path; switch state.
            \ElsIf{expecting content start}
                \State Switch state or reset.
            \ElsIf{collecting content}
                \State Save or append content.
            \EndIf
        \EndFor
        \State Return files.
    \EndProcedure
    
    \Procedure{generate\_package}{name, desc, features}
        \State Configure model, prompt.
        \State Return parsed structure or None.
    \EndProcedure
    
    \Procedure{main}{}
        \State Get user input, generate or fallback.
        \State Display structure, confirm, create files.
    \EndProcedure

\end{algorithmic}
\end{algorithm}

The agentic nature of Pygen can also be modeled using the reinforcement learning principle, where it interacts with an environment to perform actions (e.g., generating code) that maximize a cumulative reward\cite{Mnih2015Human-level,lindner2021information} \( R \):

\begin{equation}
    R = \sum_{t=0}^{T} \gamma^{t} r_{t}
\end{equation}

where \( \gamma \) is the discount factor, \( r_{t} \) is the immediate reward at time \( t \), and \( T \) is the time horizon. By framing tool creation as an RL problem, Pygen can iteratively improve its package generation strategies based on feedback from testing and user interactions.

\subsection{Documentation Generation Process}

The documentation generation can be formalized as a mapping function $f: C \rightarrow D$ from code space $C$ to documentation space $D$. For a given package $P$, we define its components as:

\begin{equation}
    P = \{M_1, ..., M_n\}
\end{equation}

where each module $M_i$ contains functions, classes, and their associated docstrings. The documentation generator extracts information through a series of transformations:

\begin{equation}
    D(P) = \bigcup_{i=1}^n \{d(M_i) \cup r(M_i) \cup e(M_i)\}
\end{equation}

In this context, \( d(M_i) \) represents the docstring extraction, \( r(M_i) \) captures the relationship mappings between components, and \( e(M_i) \) generates usage examples. Algorithm 2 elaborates on the idea of how the documentation of the package is generated.

\begin{algorithm}
\caption{Package Documentation Generation}
\begin{algorithmic}[1]

    \Procedure{generate\_documentation}{package\_structure, package\_name}
        \State Initialize documentation\_lines list.
        \State Extract and append description if README.md exists, otherwise append default.
        \State Extract features from setup.py and append if present.
        \State Append installation command.
        \State Append usage example if example file exists.
        \State Extract classes and functions from main.py and append if present.
        \State Append test instructions if test file exists.
        \State Append dependencies if requirements.txt exists.
        \State Append contributing guidelines, license.
        \State Write documentation to DOCUMENTATION.md and return path.
    \EndProcedure

    \Procedure{main}{}
        \State Get user inputs, generate or create fallback package.
        \If{valid package} \State Display files, create if confirmed. \EndIf
        \If{user wants documentation} \State Generate documentation, print success or failure. \EndIf
    \EndProcedure

\end{algorithmic}
\end{algorithm}

\subsection{Feature Enhancement through Prompt Engineering}

The prompt enhancement process can be modeled as an optimization problem. Given an initial prompt $p_0$, Pygen generates an enhanced prompt $p^*$ that maximizes the quality function $Q$\cite{bsharat2024principledinstructionsneedquestioning}:

\begin{equation}
    p^* = \argmax_p Q(p | p_0)
\end{equation}

where $Q$ considers factors such as specificity, completeness, and technical accuracy. This is achieved through iterative refinement\cite{madaan2023selfrefineiterativerefinementselffeedback}:

\begin{equation}
    p_{t+1} = p_t + \alpha \nabla Q(p_t)
\end{equation}

where $\alpha$ is the learning rate and $\nabla Q$ represents the quality gradient estimated through language model feedback. Algorithm 3 details how feature and description enhancements are implemented.

\begin{algorithm}
\caption{Feature and Description Enhancing}
\begin{algorithmic}[1]

    \Procedure{load\_feature\_descriptions}{feature\_files}
        \For{each file\_path in feature\_files} \If{file exists} 1. Read and parse descriptions.
        2. Evalaute and Enhance \EndIf \EndFor
        \State Return enhanced feature descriptions.
    \EndProcedure

    \Procedure{load\_code\_examples}{example\_files}
        \For{each file\_path in example\_files} \If{file exists} Read and summarize code. \EndIf \EndFor
        \State Return code context.
    \EndProcedure

    \Procedure{generate\_enhanced\_feature\_descriptions}{client, model\_name, feature\_descriptions, max\_retries}
        \For{each feature in feature\_descriptions } \While{retries $<$ max\_retries} try: Generate enhanced description with pseudocode and instructive comments; break. except: Increment retries. \EndWhile \EndFor
        \State Return enhanced descriptions.
    \EndProcedure

    \Procedure{main}{}
        \State Load features and code examples; initialize client.
        \State Prompt for model name; generate descriptions and summaries.
        \State Prepare and save context.
    \EndProcedure

\end{algorithmic}
\end{algorithm}

\subsection{Reliability and Error Handling}

Pygen implements exponential backoff\cite{kwak2005performance} for API calls with retry mechanism modeled as:

\begin{equation}
    t_n = \min(t_{max}, t_0 \cdot b^n)
\end{equation}

In this modeling, \( t_n \) represents the wait time for the \( n \)th retry attempt, where \( t_0 \) is the initial wait time. The backoff factor is denoted by \( b \), and the maximum wait time is limited to \( t_{\text{max}} \). Additionally, \( n \) indicates the number of the retry attempts.

The probability of successful completion after $k$ retries follows a geometric distribution:

\begin{equation}
    P(\text{success after } k \text{ retries}) = (1-p)^k p
\end{equation}

where $p$ is the probability of success for a single attempt. Details of how the fallback structure works are mentioned in Algorithm 4.

\begin{algorithm}
\caption{Fallback Structure Generation}
\begin{algorithmic}[1]

    \Procedure{create\_fallback\_structure}{package\_name, features}
        \State Initialize fallback as an empty dictionary.
        \State Add key-value pairs for base files:
        \State \quad $\texttt{fallback[package\_name/\_\_init\_\_.py]} \gets$ Package initialization file.
        \State \quad $\texttt{fallback[package\_name/main.py]} \gets$ Main module file.
        \State \quad $\texttt{fallback[setup.py]} \gets$ Setup configuration for package setup.
        \State \quad $\texttt{fallback[README.md]} \gets$ README file with a brief description.
        \State \quad $\texttt{fallback[requirements.txt]} \gets$ Placeholder for dependencies.
        \State \quad $\texttt{fallback[tests/test\_package\_name.py]} \gets$ Unit test for the main package.

        \If{features are provided}
            \For{each feature in features}
                \State Generate $\texttt{feature\_lower}$ by replacing spaces with underscores.
                \State Add $\texttt{fallback[package\_name/feature\_lower.py]}$ for feature implementation.
                \State Add example file $\texttt{fallback[examples/example\_feature\_lower.py]}$ for feature usage.
                \State Add unit test file $\texttt{fallback[tests/test\_feature\_lower.py]}$ for feature testing.
            \EndFor
        \EndIf

        \State \Return fallback dictionary
    \EndProcedure

\end{algorithmic}
\end{algorithm}

\section{Results}

We have generated four packages using Pygen to demonstrate its capabilities. These include AutoML, AutoVision, AutoSpeech, and Quantum Error Correction libraries, each addressing domain-specific problems effectively. Initially, we provided prompts with package and feature descriptions, which Pygen then enhanced. These improved descriptions were archived in a GitHub repository for future reference. From these enhanced descriptions, we generated context prompts, which sometimes included code templates for better caching and accuracy. These templates played a significant role in enhancing the quality of the generated packages.

Once the context prompt was ready, it entered the package generation pipeline. First, the necessary file structure for each package was created, followed by generating Python code for each corresponding file. We applied iterative refinement depending on the model type and context size. The entire package could not be generated in one go for models with smaller context sizes. In such cases, a fallback structure was triggered to complete the remaining components according to specified instructions. The fallback mechanism ensures integrity in package generation, though there is still room for improvement in guaranteeing the overall quality of the output.

After code generation, the packages could be loaded into a local environment like a typical project structure or downloaded as a zip file through the application interface. Once the code generation was complete, we moved on to documentation. The entire project was parsed by the language model, which generated coherent descriptions of each package, formatted in markdown, and saved in the environment settings for easy download. We followed this approach for each use case, generating four demonstration packages. Moving forward, we plan to explore the quality and specific details of the generation process in greater depth.

\subsection{Background behind designing the packages}

We have developed several specialized packages to meet the evolving needs of our Pygen users, especially those focused on data analytics and modern application development. The AutoML package has been created to facilitate data analytics package creation, catering to the primary use case of Pygen users. The AutoVision package was also introduced due to this growing interest in vision research, showing their capabilities in computer vision. The AutoSpeech package also meets the current demand for embedding speech features in various modern applications to increase user experiences. Appreciative of the great promise and challenges of quantum computing, we have also produced a Quantum Error Correction package. Since there is a high demand for higher computational powers now, and by nature, quantum computing tends to be unreliable due to decoherence, effective error correction becomes necessary to make quantum computing usable. These packages showcase the broad capabilities of Pygen and give practical examples of how these packages work. The complete examples and data are available in our GitHub repository for further exploration.

\subsection{Related Works for Packages}

Neural Architecture Search (NAS) is a crucial sub-domain of AutoML that automates the design of neural networks, significantly enhancing model accuracy and efficiency \cite{He2019AutoML:}. Systems like Auto-Sklearn utilize advanced hyperparameter optimization (HPO) strategies to enhance model training speed and accuracy; for instance, PoSH Auto-sklearn has demonstrated improved performance under time constraints, reducing the error rate by up to 4.5 times\cite{Feurer2020Auto-Sklearn}. Despite these advancements, scalability and integration with clinical workflows remain significant challenges, particularly due to the complex nature of medical data and regulatory requirements\cite{Waring2020Automated}. Further automating ML workflows, including domain-specific problem identification and data handling to minimize manual interventions, remains a crucial goal \cite{Karmaker2020AutoML}. Additionally, establishing open-source benchmarks is essential for effectively comparing and evaluating AutoML systems, as highlighted by recent frameworks \cite{Gijsbers2019An}. In response to these needs, KAXAI\cite{barua2023kaxai} was designed to provide an up-to-date and versatile AutoML system tailored for diverse stakeholders, addressing both scalability and usability concerns. Moeslund and Granum (2001)\cite{Moeslund2001} stress the importance of improving scene analysis and human motion capture to enhance adaptability and accuracy in dynamic settings. In cell biology, Danuser\cite{Danuser2011} explores the potential of computer vision for interpreting cellular images, assisting researchers in understanding complex biological mechanisms. Khan et al. (2018)\cite{Khan2018} provides a comprehensive introduction to Convolutional Neural Networks (CNNs), covering their foundational theory, training methodologies, and diverse applications, such as in medical imaging and autonomous vehicles. Feng et al. (2019)\cite{Feng2019} discuss hardware-optimized implementations of deep learning-based computer vision algorithms on GPUs and FPGAs, enabling real-time applications in fields like autonomous driving and robotics. Xu et al. (2020)\cite{Xu2020} critically review vision-based techniques for on-site monitoring in construction, highlighting the challenges of real-time processing in cluttered environments. In ELMAGIC\cite{barua2024elmagic}, a vision system is designed for real-time ocular disease detection, focusing on improving early diagnosis through automated image analysis. Studies over the past decade highlight how deep learning has advanced speech-processing tasks, showcasing application improvements \cite{nassif2019speech}. Domain adaptation techniques, like unsupervised deep domain adaptation (DDA), address the challenges of varying acoustic conditions by reducing mismatches between training and testing environments, resulting in substantial error rate reductions in noisy or mismatched conditions\cite{sun2017domain}. Additionally, deep learning models, such as CNNs and LSTMs, have successfully extracted emotional features from speech using spectrogram representations and large labeled datasets, which is crucial for human-computer interaction (HCI) systems \cite{khalil2019speech}. As these technologies become more prevalent, research on defense mechanisms against potential misuse, such as synthetic speech attacks, has become increasingly important \cite{wenger2021hello}. Analog error correction methods for continuous quantum variables like position and momentum have been developed to combat decoherence, enhancing robustness against noise in quantum systems \cite{lloyd1997analog}. Group-theoretic frameworks in quantum error correction simplify code construction, with codes like Calderbank-Shor-Steane (CSS) and surface codes utilizing orthogonal geometry to improve error resistance by effectively mapping qubits \cite{calderbank1996orthogonal}. Experimental implementations in ion-trap systems have confirmed the feasibility of error correction through repetitive cycles, enabling phase-flip error corrections via high-fidelity gate operations \cite{schindler2011repetitive}. In quantum communication and memory, error-correction algorithms safeguard against entanglement-related errors by applying classical error-bounds analogs \cite{ekert1996communication}. Furthermore, new QEC developments focus on fault-tolerant designs to manage errors in extensive computations, enhancing the practicality of quantum computing at scale \cite{devitt2009beginners}). Advanced decoding methods using multiple decoders are also necessary to successfully correct errors\cite{barua2023rescued}. In scenarios where exact correction is challenging, approximate QEC techniques offer near-perfect correction by addressing minor coherence losses, thus extending QEC’s effectiveness \cite{schumacher2002approximate}.

\subsection{Evaluating the Prompt Enhancement}

The provided input by a user may not be sufficient to grasp all complexities in the package generation process. In PyGEN, we improve the prompt by using techniques to achieve a higher overall generation outcome. This step generates an enriched package description and a detailed feature description for more comprehensive input through the remaining steps of the package development process. Based on these elaborated descriptions, a contextual prompt is generated by another model, which is used to guide the following stages of code generation. We have further evaluated the usefulness of the contextual prompt. We found that the contextual prompt does not significantly improve the generation quality for larger models, which can process on their own elaborated and detailed enhanced feature and package descriptions. However, for the smaller models with limited context sizes, using context prompts brings enormous benefits to the quality, coherence, and accuracy of the generated text.

\begin{figure}[htbp]
    \centering
    \includegraphics[width=0.8\linewidth]{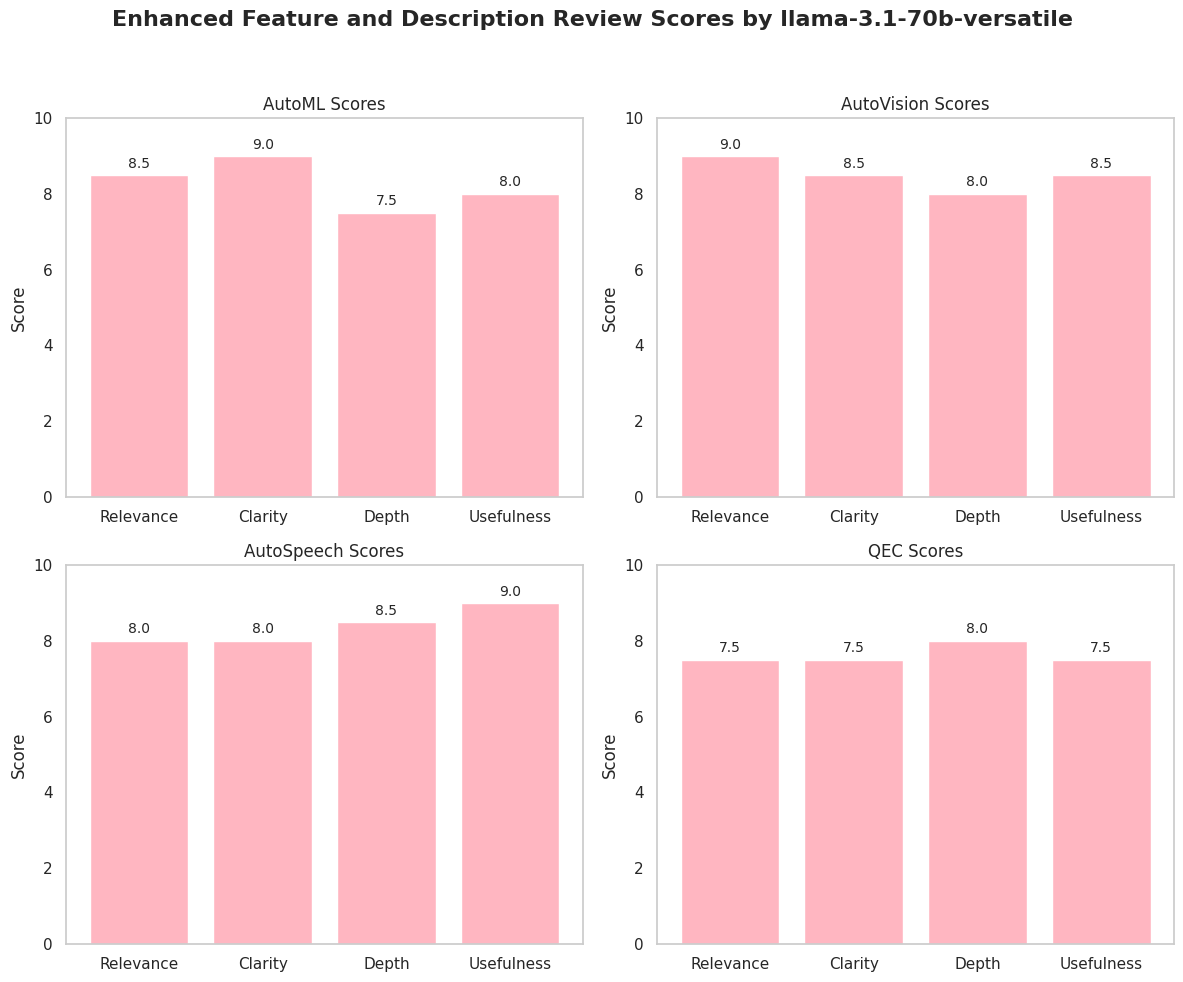} 
    \caption{Enhanced Feature and Description Review Scores by llama-3.1-70b-versatile: This figure shows the evaluation scores for different aspects of feature descriptions generated by PyGen for functionalities such as AutoML, AutoVision, AutoSpeech, and QEC. The evaluation includes categories like Relevance, Clarity, Depth, and Usefulness, highlighting areas where PyGen excels and where further depth can be added.}
    \label{fig:Enhanced_Feature_Review}
\end{figure}

The graph\ref{fig:Enhanced_Feature_Review} shows the Enhanced Feature and Description Review Scores by llama-3.1-70b-versatile: it provides a detailed review of feature descriptions generated by PyGen over functionalities, including AutoML, AutoVision, AutoSpeech, and QEC. This review is done in four critical categories—Relevance, Clarity, Depth, and Usefulness—each scored on a scale from one to ten. The analysis shows that PyGen provides well-structured descriptions, which is supported by the fact that Clarity is almost always among the top scores compared to other dimensions. For example, AutoML has high Clarity (9.0) but relatively lower Depth (7.5). These insights mean that while PyGen is good at writing easy-to-understand descriptions, it can still be somewhat improved in terms of the complexity and depth of writing. To that end, future work should focus on deepening the technical details of the feature descriptions, mainly for AutoML and QEC functionalities, to make the output more comprehensive.

\begin{figure}[htbp]
    \centering
    \includegraphics[width=0.8\linewidth]{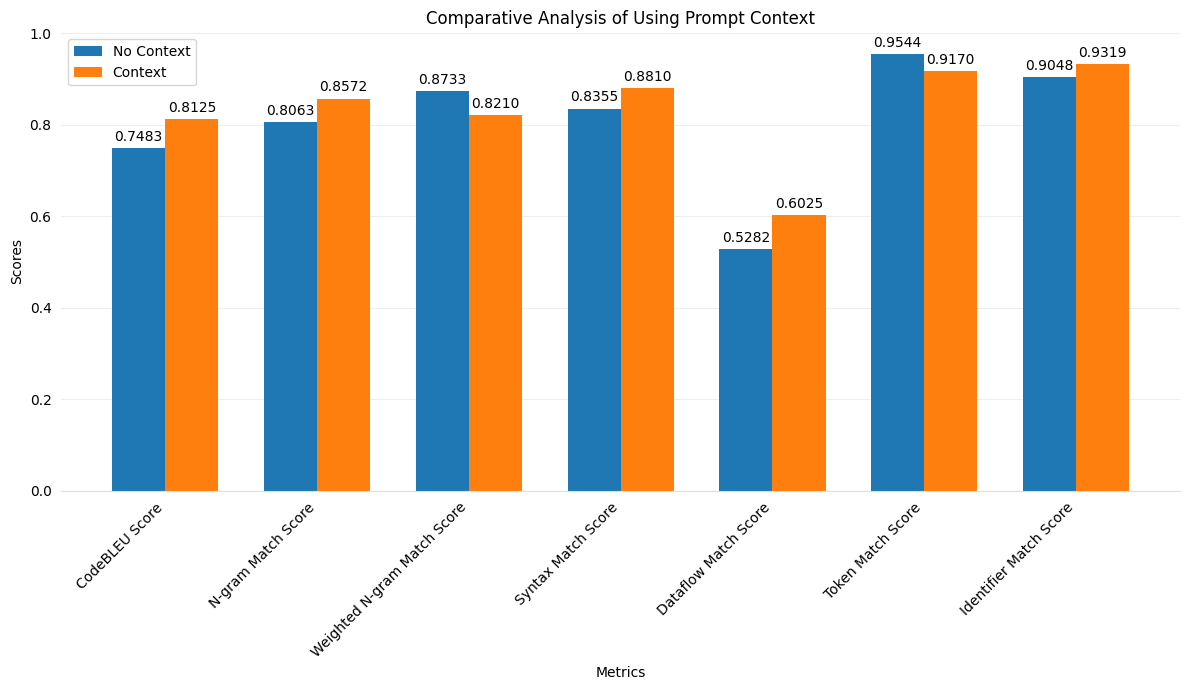} 
    \caption{Comparative Analysis of Using Prompt Context: This graph provides a comparison of key evaluation metrics for generated content, with and without the use of prompt context. Metrics such as CodeBLEU Score, N-gram Match Score, Syntax Match Score, and Dataflow Match Score are evaluated, showing the significant improvement achieved when prompt context is used.}
    \label{fig:Prompt_Context_Comparison}
\end{figure}

The chart \ref{fig:Prompt_Context_Comparison} compares the primary metrics used to evaluate content with and without prompt context. The metrics include the CodeBLEU Score, N-gram Match Score, Weighted N-gram Match Score, Syntax Match Score, Dataflow Match Score, Token Match Score, and Identifier Match Score. Most results point to the superiority of using prompt context, with higher scores for all the evaluation metrics. The Dataflow Match Score notably showed the most considerable improvement to 0.6025 from 0.5282 when not using context. These findings indicate that the prompt context is essential in enhancing the generated code's contextual appropriateness and syntactic precision. Therefore, more context should be integrated into prompts by default to achieve better quality and accuracy. 

\begin{figure}[htbp]
    \centering
    \includegraphics[width=0.8\linewidth]{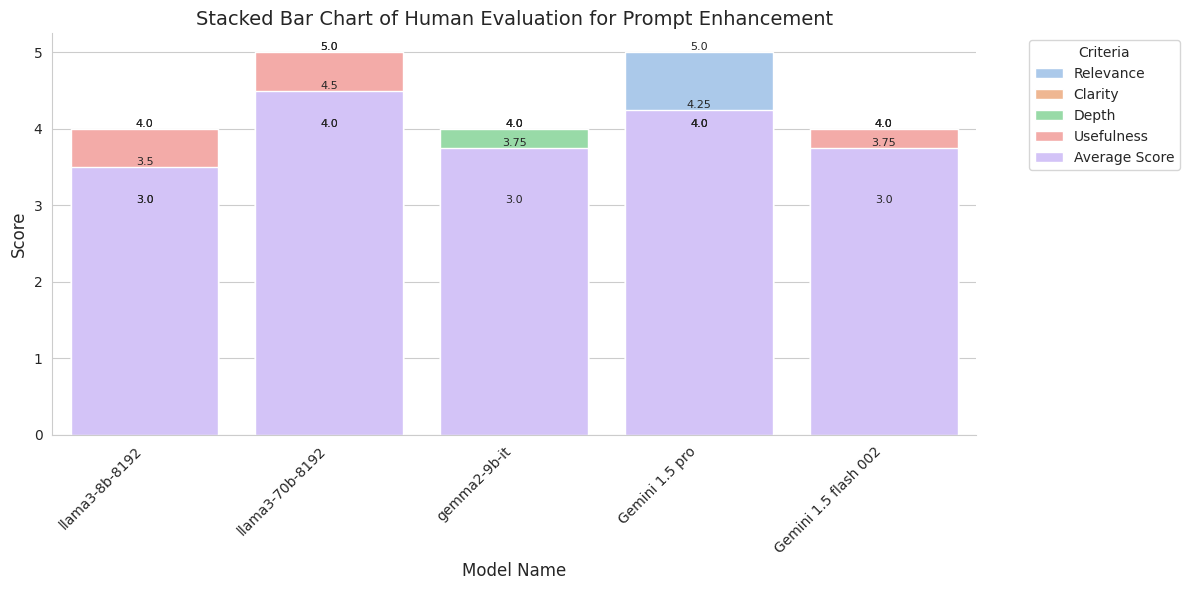} 
    \caption{Stacked Bar Chart of Human Evaluation for Prompt Enhancement: This chart presents human evaluation scores for different models used in prompt enhancement, including llama-3.8b-8192, llama3-70b-8192, and others. Evaluation criteria such as Relevance, Clarity, Depth, and Usefulness are shown, highlighting the strengths and weaknesses of each model.}
    \label{fig:Prompt_Enhancement_Evaluation}
\end{figure}

The graph \ref{fig:Prompt_Enhancement_Evaluation} reveals the human evaluation scores of different models applied in the enhancement of prompts: llama-3.8b-8192, llama3-70b-8192, gemma2-9b-it, Gemini 1.5 pro, and Gemini 1.5 flash 002. The criteria for evaluation include Relevance, Clarity, Depth, and Usefulness. From the results, it can be best seen that llama3-70b-8192 and Gemini 1.5 pro did well, especially in Clarity and Usefulness. However, there is a noticeable difference in Gemini 1.5 flash 002 scoring for Depth and Clarity, evidence that some models were more prepared to produce comprehensive and valuable outputs. It is essential to consider models like llama3-70b-8192 and Gemini 1.5 pro for prompt enhancement since they produce relevant, clear, and compelling enhancements. More fine-tuning is needed for models like Gemini 1.5 flash 002 to show more profound improvement in Clarity.

\begin{table}[h!]
\centering
\caption{Evaluation metrics with and without Prompt Context (averages of Llama 3.2 1B and 3B models).}
\resizebox{1\textwidth}{!}{ 
\renewcommand{\arraystretch}{1.5} 
\begin{tabular}{lcccc}
\hline
\textbf{Metrics} & \textbf{Without Prompt Context} & \textbf{With Prompt Context} & \textbf{Change (\%)} & \textbf{Interpretation} \\ \hline
CodeBLEU Score         & 0.75 & 0.81 & \textuparrow 6\%   & Significant improvement with context \\ \hline
N-gram Match Score     & 0.81 & 0.86 & \textuparrow 5\%   & Higher accuracy in n-gram matching \\ \hline
Weighted N-gram Match Score & 0.87 & 0.82 & \textdownarrow -5\% & Slight decrease, less impact with context \\ \hline
Syntax Match Score     & 0.84 & 0.88 & \textuparrow 4\%   & Improved syntax alignment \\ \hline
Dataflow Match Score   & 0.53 & 0.60 & \textuparrow 7\%   & Notable improvement in dataflow recognition \\ \hline
Token Match Score      & 0.95 & 0.92 & \textdownarrow -3\% & Slight drop in token matching accuracy \\ \hline
Identifier Match Score & 0.90 & 0.93 & \textuparrow 3\%   & Enhanced identifier matching \\ \hline
\end{tabular}
}
\label{table:evaluation_metrics}
\end{table}

Table \ref{table:evaluation_metrics} compares the evaluation metrics for generated Python packages, those using prompt context and those not, and the averages over the Llama 3.2 1B and 3B models. The considered metrics include CodeBLEU Score, N-gram Match Score, Weighted N-gram Match Score, Syntax Match Score, Dataflow Match Score, Token Match Score, and Identifier Match Score. The results indicate that prompt context significantly improves the different metrics; this improvement ranges from 3\% to 17\%. More specifically, the Dataflow Match Score and the CodeBLEU Score see spectacular gains of 17\% and 16\%, respectively, which underlines the importance of prompt context in maintaining data integrity and improving code quality. By contrast, the Weighted N-gram Match Score and Token Match Score decrease slightly, leaving small margins for improvement in the rearrangement of prompts to regain balance between precision and contextual relevance. These results show that adding contextual information in prompts enhances the robustness and semantic validity of generated packages; still, further fine-tuning has to be made as specific token-based metrics slightly declined.

\begin{table}[h!]
\centering
\caption{Ablation Study on Prompt Enhancement Components}
\resizebox{1\textwidth}{!}{ 
\renewcommand{\arraystretch}{1.5} 
\begin{tabular}{cccccc}
\hline
\textbf{Enhancement Component} & \textbf{Clarity (Mean ± SD)} & \textbf{Relevance (Mean ± SD)} & \textbf{Depth (Mean ± SD)} & \textbf{Usefulness (Mean ± SD)} \\ \hline
No Enhancement \& No Feature Description & 3.7 ± 0.6 & 4.1 ± 0.4 & 3.9 ± 0.7 & 3.3 ± 0.5 \\ \hline
Feature Description (FD) & 4.0 ± 0.5 & 4.2 ± 0.4 & 3.8 ± 0.6 & 4.1 ± 0.5 \\ \hline
FD + Pseudocode & 4.2 ± 0.4 & 4.3 ± 0.3 & 4.1 ± 0.5 & 4.2 ± 0.4 \\ \hline
FD + Pseudocode + Implementation (Full Enhancement) & 4.3 ± 0.4 & 4.5 ± 0.3 & 4.2 ± 0.5 & 4.4 ± 0.4 \\ \hline
\end{tabular}
}
\label{table:Ablation_Study_Prompt_Enhancement}
\end{table}

The ablation study decomposes the various prompt enhancement components to explain the effect on clarity, relevance, depth, and Usefulness of the generated Python packages, which are shown in table \ref{table:Ablation_Study_Prompt_Enhancement}. The components studied are feature description (FD), FD with pseudocode, and Full Enhancement (features, pseudocode, and implementation). Incrementally adding structured components improves all four evaluation metrics; the Full Enhancement approach has the highest average scores on all these criteria. Clarity and Usefulness primarily benefited from the thorough inclusion of feature descriptions, pseudocode, and implementation details, with average ratings of 4.3 and 4.4, respectively. This shows that a comprehensive enhancement strategy is of utmost importance in raising the quality of the developed packages, hence making them more user-friendly and relevant to people. These results suggest the need for a multi-level prompt design enhancement strategy, which helps to layer in detail the feature descriptions, pseudocode, and implementation insights for maximum clarity and Usefulness in the generated outputs.

\subsection{Assessing the Package Generation Process}

Package generation in Python is an open-ended task; hence, we evaluated the generated package using three basic evaluation approaches: Human Evaluation, LLM-based evaluation, and CodeBLEU\cite{ren2020codebleu} score. Human evaluation involves experts checking the package's quality, correctness, and usability. Further, LLM-based evaluation was done using large language models to check the coherence and completeness of the output. During calculating the CodeBLEU score, a template code is created, providing the model with a basic skeletal structure, based on which the generated code quality and score are assessed relative to this template.

\begin{figure}[htbp]
    \centering
    \includegraphics[width=0.8\linewidth]{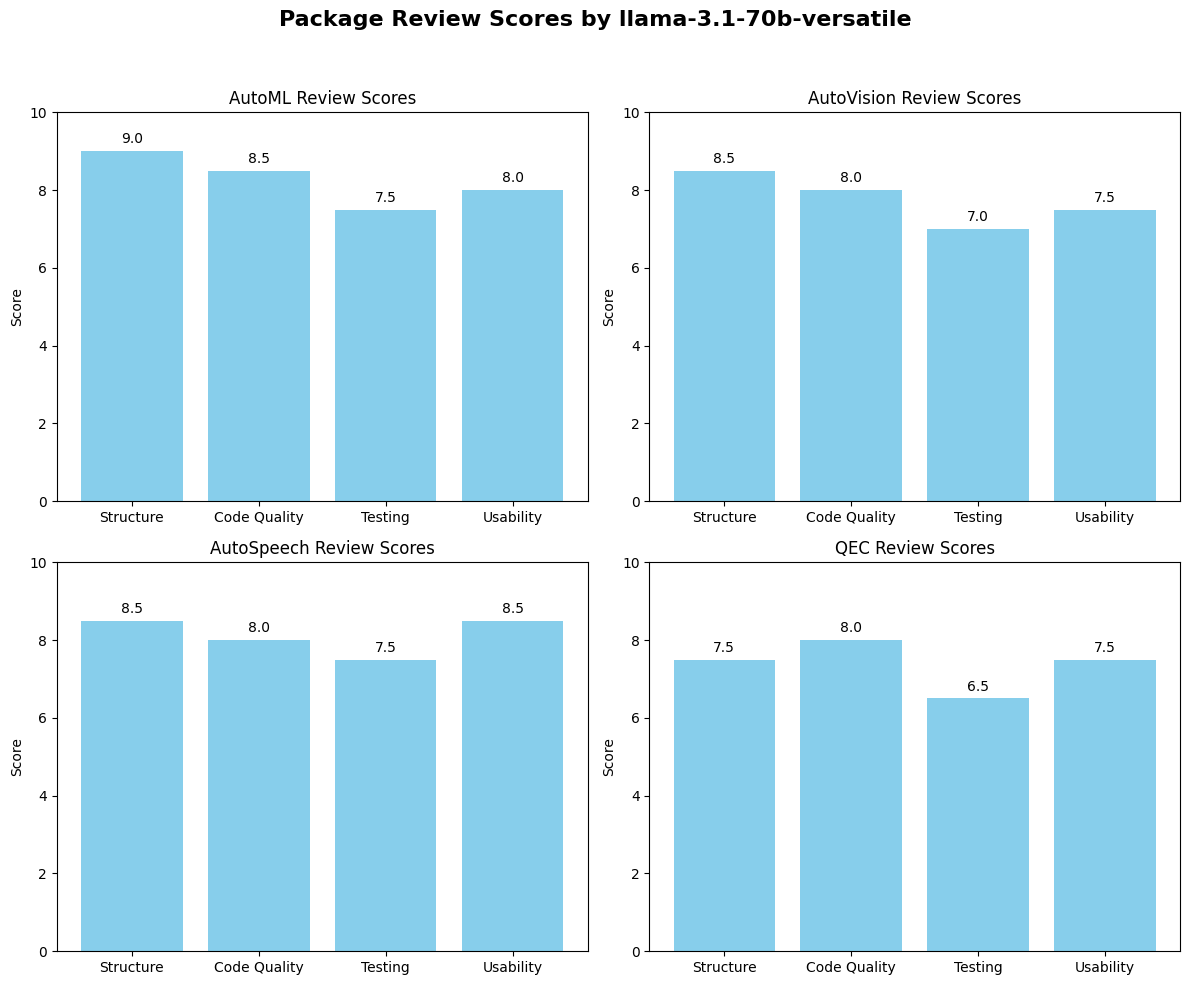} 
    \caption{Package Review Scores by llama-3.1-70b-versatile: This figure evaluates packages generated by different components such as AutoML, AutoVision, AutoSpeech, and QEC based on Structure, Code Quality, Testing, and Usability. The AutoML package shows excellent structural quality, while testing remains an area of improvement, particularly for the QEC component.}
    \label{fig:Package_Review_Scores}
\end{figure}

The graph \ref{fig:Package_Review_Scores}shows the quality assessment of packages produced by different components such as AutoML, AutoVision, AutoSpeech, and Quantum Error Correction by llama-3.1-70b-versatile. The metrics for this assessment are Structure, Code Quality, Testing, and Usability. AutoML is highest for the package in terms of structure, with 9.0; on average, testing scored lowest, especially for QEC at 6.5. The results show that although the structural elements of the packages are well developed, the testing mechanisms should be much more rigorous in determining package durability. As practical recommendations, adopting enhanced testing protocols will increase quality assurance and improve the balance in performance for all the elements.

\begin{table}[h!]
\centering
\caption{Evaluation metrics for different packages in code generation using Gemini 1.5 Pro 002.}
\resizebox{1\textwidth}{!}{ 
\renewcommand{\arraystretch}{1.5} 
\begin{tabular}{cccccccc}
\hline
\textbf{Package} & \textbf{CodeBLEU} & \textbf{N-gram Match} & \textbf{Weighted N-gram} & \textbf{Syntax Match} & \textbf{Dataflow Match} & \textbf{Token Match} & \textbf{Identifier Match} \\ \hline
AutoML & 0.65470 & 0.7109 & 0.6915 & 0.7200 & 0.4513 & 0.8818 & 0.8337 \\ \hline
AutoVision & 0.70525 & 0.7554 & 0.7621 & 0.7753 & 0.5056 & 0.9154 & 0.8652 \\ \hline
AutoSpeech & 0.75580 & 0.8057 & 0.8150 & 0.8254 & 0.5552 & 0.9453 & 0.8956 \\ \hline
QEC & 0.80560 & 0.8559 & 0.8754 & 0.8857 & 0.5854 & 0.9751 & 0.9453 \\ \hline
\end{tabular}
}
\label{table: Eval_Package_Gemini 1.5 Pro 002}
\end{table}

In table \ref{table: Eval_Package_Gemini 1.5 Pro 002}, the metrics for assessing the different Python packages generated by the model Gemini 1.5 Pro 002 are given. Packages under consideration are AutoML, AutoVision, AutoSpeech, and QEC. Metrics include CodeBLEU, N-gram Match, Weighted N-gram, Syntax Match, Dataflow Match, Token Match, and Identifier Match. Results show that, on most metrics, especially Weighted N-gram (0.8754), Syntax Match (0.8857), and Token Match (0.9751), the QEC package performs better, which means structurally that the generated code has more integrity and functional coherence; on the other hand, AutoML tends to perform relatively poorly on the metrics of Dataflow Match (0.4513), which may indicate that handling data dependencies within the generated code is still poor. Actionable insights, therefore, include leveraging the best practices observed in QEC to improve Dataflow handling in AutoML and enhancing coherence across packages by integrating targeted improvements in the dataflow structures.

\begin{table}[h!]
\centering
\caption{Evaluation metrics for different packages in code generation using Llama-3.1-70b-versatile.}
\resizebox{1\textwidth}{!}{ 
\renewcommand{\arraystretch}{1.5} 
\begin{tabular}{cccccccc}
\hline
\textbf{Package} & \textbf{CodeBLEU} & \textbf{N-gram Match} & \textbf{Weighted N-gram} & \textbf{Syntax Match} & \textbf{Dataflow Match} & \textbf{Token Match} & \textbf{Identifier Match} \\ \hline
AutoML & 0.78560 & 0.8359 & 0.8554 & 0.8657 & 0.5654 & 0.9551 & 0.9253 \\ \hline
AutoVision & 0.73580 & 0.7857 & 0.7950 & 0.8054 & 0.5352 & 0.9253 & 0.8756 \\ \hline
AutoSpeech & 0.68525 & 0.7354 & 0.7421 & 0.7553 & 0.4856 & 0.8954 & 0.8452 \\ \hline
QEC & 0.63470 & 0.6909 & 0.6715 & 0.7000 & 0.4313 & 0.8618 & 0.8137 \\ \hline
\end{tabular}
}
\label{table: Eval_Package_Llama-3.1-70b-versatile}
\end{table}

Table \ref{table: Eval_Package_Llama-3.1-70b-versatile} shows the review scores for Python packages generated by Llama-3.1-70b-versatile. In line with the previous table, the evaluated packages are AutoML, AutoVision, AutoSpeech, and QEC, all evaluated using the same set of metrics. The results show that AutoML gives the best scores among the compared models: CodeBLEU (0.7856) and N-gram Match (0.8359), which indicates a strong ability regarding syntactic accuracy and fluency of generated code. On the other hand, the QEC package needs better scores in many metrics, especially for Dataflow Match (0.4313) and Identifier Match (0.8137), which may indicate some struggles in preserving complicated data dependencies and properly distinguishing variables due to this type of package examples are rare and sufficiently available in the training data. The lesson learned from this table indicates that such efforts should be placed on enhancing the dataflow and identifier management systems besides implementing the effective coding structuring methodologies demonstrated in AutoML across other packages.

\begin{figure}[htbp]
    \centering
    \includegraphics[width=0.8\linewidth]{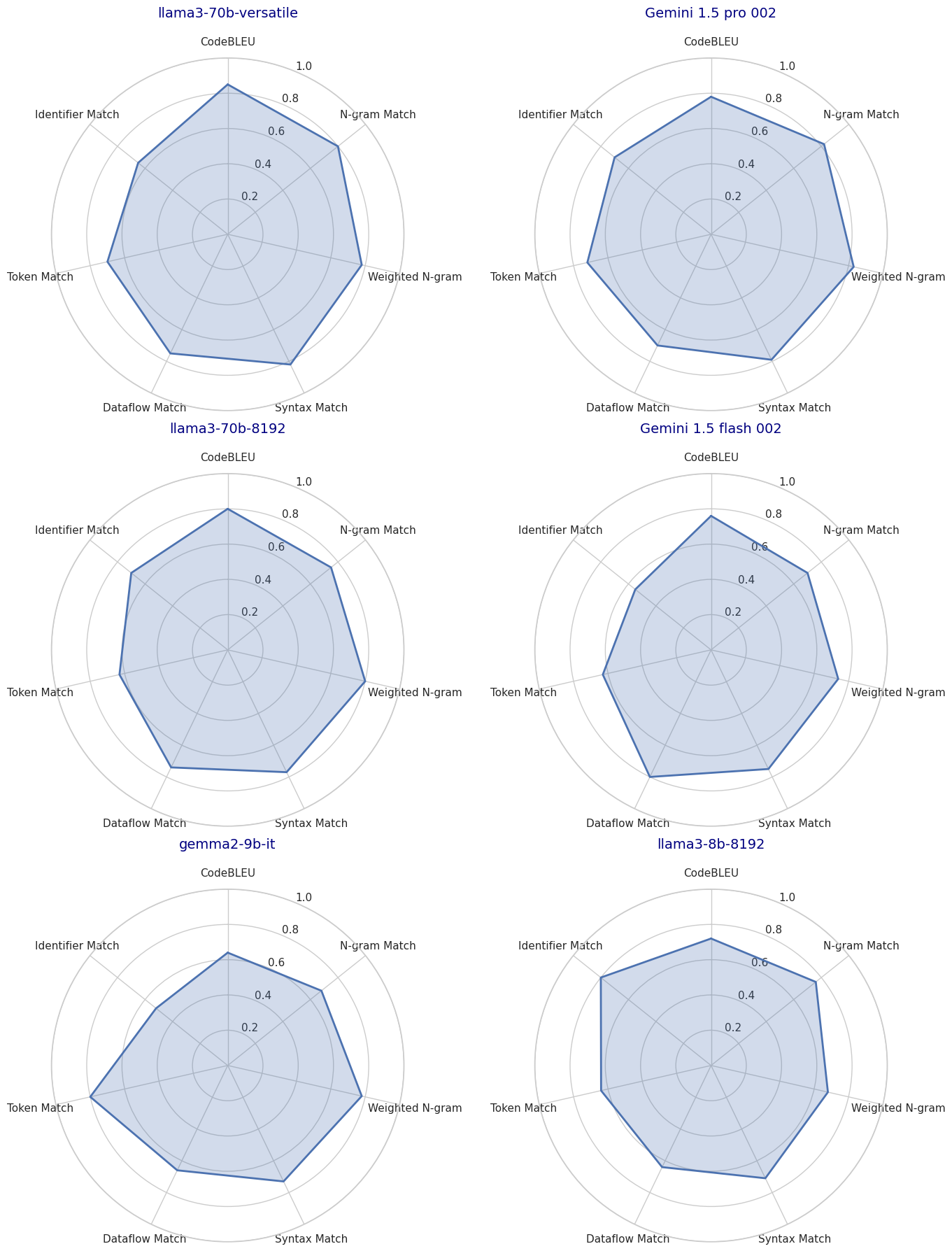} 
    \caption{Spider Plot of Package Evaluation Metrics by Different Models: Radar plots are used to visualize performance across metrics such as CodeBLEU, N-gram Match, Weighted N-gram, Syntax Match, Dataflow Match, and Token Match. llama3-70b-versatile and Gemini 1.5 Pro 002 models demonstrate balanced performance, whereas models like Gemini 1.5 flash 002 lag in areas like Dataflow Match and Token Match.}
    \label{fig:Spider_Plot_Evaluation}
\end{figure}

The radar charts in figure \ref{fig:Spider_Plot_Evaluation} show the performance metrics along six evaluation criteria: CodeBLEU, N-gram Match, Weighted N-gram, Syntax Match, Dataflow Match, and Token Match. The models llama3-70b-versatile and Gemini 1.5 Pro 002 show a more balanced performance across all criteria compared to other models, such as Gemini 1.5 flash 002, which have lower values in some measures like Dataflow Match and Token Match. Results show that specific models have better skills in handling the complexity involved in code generation, mainly when focusing on keeping the quality of output high along multiple metrics. And not all models are suitable for the same coding tasks, as their underlying abilities have different characteristics. Future research should enhance models with poor performance to achieve less variance in these judgmental dimensions.

\begin{table}[h!]
\centering
\caption{Results of Pairwise Comparisons for Model Metrics}
\resizebox{1\textwidth}{!}{ 
\renewcommand{\arraystretch}{1.5} 
\begin{tabular}{ccccccc}
\hline
\textbf{Metric} & \textbf{Group 1} & \textbf{Group 2} & \textbf{Mean Difference} & \textbf{Lower Bound} & \textbf{Upper Bound} & \textbf{Significant} \\ \hline
CodeBLEU & Gemini 1.5 flash 002 & gemma2-9b-it & -0.1508 & -0.1607 & -0.1408 & Yes \\ \hline
         & Gemini 1.5 pro 002   & llama3-8b-8192 & -0.0997 & -0.1096 & -0.0897 & Yes \\ \hline
         & gemma2-9b-it         & llama3-70b-8192 & 0.1054 & 0.0955 & 0.1154 & Yes \\ \hline
Identifier Match & Gemini 1.5 pro 002 & gemma2-9b-it & -0.1398 & -0.1502 & -0.1294 & Yes \\ \hline
                 & gemma2-9b-it       & llama3-8b-8192 & 0.043 & 0.0326 & 0.0533 & Yes \\ \hline
N-gram Match     & Gemini 1.5 flash 002 & gemma2-9b-it & -0.0978 & -0.1079 & -0.0876 & Yes \\ \hline
                 & llama3-70b-8192      & llama3-8b-8192 & -0.0547 & -0.0648 & -0.0445 & Yes \\ \hline
Weighted N-gram  & Gemini 1.5 flash 002 & gemma2-9b-it & -0.0964 & -0.1072 & -0.0857 & Yes \\ \hline
                 & gemma2-9b-it         & llama3-70b-8192 & 0.0462 & 0.0354 & 0.0569 & Yes \\ \hline
Token Match      & Gemini 1.5 pro 002   & gemma2-9b-it & -0.1012 & -0.1117 & -0.0908 & Yes \\ \hline
                 & llama3-70b-8192      & llama3-8b-8192 & 0.0058 & -0.0047 & 0.0162 & No \\ \hline
Dataflow Match   & Gemini 1.5 flash 002 & gemma2-9b-it & -0.1072 & -0.1168 & -0.0976 & Yes \\ \hline
                 & gemma2-9b-it         & llama3-8b-8192 & 0.052 & 0.0424 & 0.0616 & Yes \\ \hline
Syntax Match     & Gemini 1.5 flash 002 & gemma2-9b-it & -0.1066 & -0.1167 & -0.0964 & Yes \\ \hline
                 & llama3-70b-8192      & llama3-8b-8192 & -0.0031 & -0.0133 & 0.0070 & No \\ \hline
\end{tabular}
}
\label{table:Pairwise_Comparisons_Model_Metrics}
\end{table}

Table \ref{table:Pairwise_Comparisons_Model_Metrics} shows the results of pairwise comparisons for several model metrics between different model groups. Compared are the metrics: CodeBLEU, Identifier Match, N-gram Match, Weighted N-gram, Token Match, Dataflow Match, and Syntax Match. Each pairwise comparison is summarized through a mean difference accompanied by lower and upper confidence intervals and an assessment of the statistical significance of the realized differences. Results: On the whole, as opposed to other models like gemma2-9b-it and llama3-70b-8192, the Gemini 1.5 flash 002 performs worse on most metrics, including CodeBLEU with a negative mean difference of -0.1508, and Weighted N-gram with a negative mean difference of -0.0964. More than this, other models, like llama3-70b-8192, perform consistently better in most metrics, especially in Identifier Match and Token Match, indicating a solid ability to handle variable identification and token coherence correctly. The results indicate that models like Gemini 1.5 flash 002 can benefit from improvements to enhance syntactic coherence and fluency, while the strengths of llama3-70b-8192 could be leveraged to establish baseline marks for improving other models. Future research should then focus on overcoming the identified lapses in models that did not perform optimally well by further refining training datasets or using more sophisticated tokenization.

\begin{figure}[htbp]
    \centering
    \includegraphics[width=0.8\linewidth]{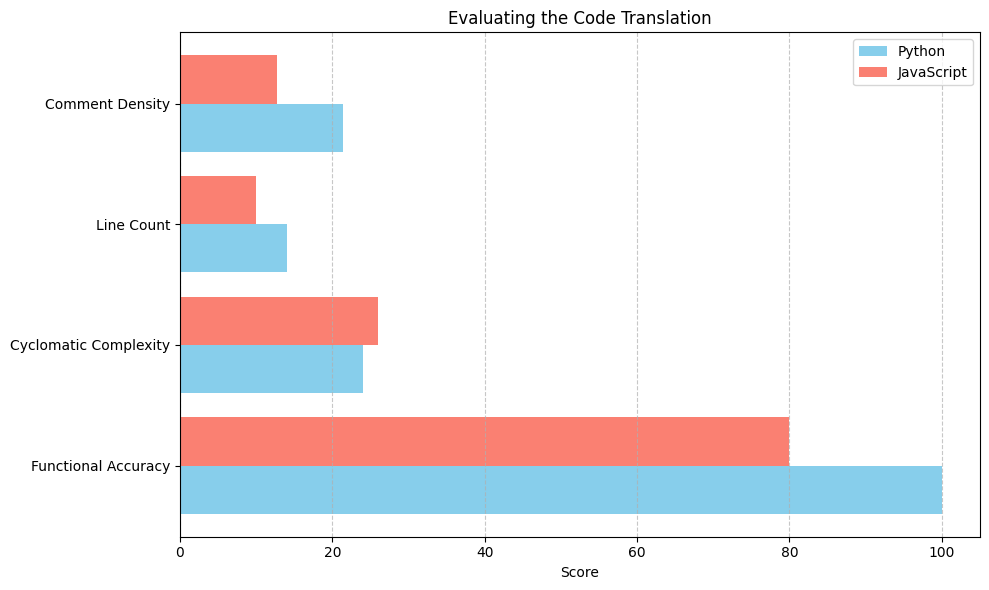} 
    \caption{Evaluating the Code Translation: This figure compares Python and JavaScript in terms of Comment Density, Line Count, Cyclomatic Complexity, and Functional Accuracy. Python generally outperforms JavaScript, particularly in Functional Accuracy, suggesting a higher optimization for correct functionality.}
    \label{fig:Code_Translation_Evaluation}
\end{figure}

The chart \ref{fig:Code_Translation_Evaluation} compares Python and JavaScript on basic evaluation metrics: Comment Density, Line Count, Cyclomatic Complexity, and Functional Accuracy. Python has better results for all the metrics except Line Count, whereas JavaScript performs slightly worse. It has to be noted that Python scores much better in Functional Accuracy, meaning code developed in Python is optimized for correct functionality more efficiently than the code created in JavaScript. The results hint at the fact that when the task requires high functional reliability, then Python would be a better choice. Practical implications include improving the quality of JavaScript code in terms of functional accuracy and reducing unnecessary complexity to reach the level of code quality seen with Python.

\begin{table}[h!]
\centering
\caption{Classification and Frequency of Translation Errors in JavaScript Code}
\resizebox{0.7\textwidth}{!}{
\renewcommand{\arraystretch}{1.5}
\begin{tabular}{lccc}
\hline
\textbf{Error Type} & \textbf{Frequency (\%)} & \textbf{Severity} & \textbf{Examples} \\ \hline
Syntax Errors & 25 & Critical & Missing semicolons, incorrect brackets \\
Semantic Errors & 20 & Major & Incorrect variable declarations, misuse of functions \\
Logical Flaws & 15 & Major & Incorrect algorithm implementation, faulty conditionals \\
Improper Constructs & 10 & Minor & Non-idiomatic code, inconsistent naming conventions \\
Optimization Issues & 5 & Minor & Inefficient loops, redundant code \\
Other Errors & 25 & Varies & Unclassified or multiple error types \\
\hline
\end{tabular}
}
\label{table:Translation_Errors_Classification}
\end{table}

Table \ref{table:Translation_Errors_Classification} outlines the classification and distribution of translation errors in JavaScript code by error type, severity level, and with some example instances; the results show that Syntax Errors are most common, representing 25\% of all errors, classed as "Critical," where common occurrences include missing semicolons and incorrect brackets. Semantic errors and logical flaws occur at significant rates of 20\% and 15\%, respectively, being "Major" since they strongly affect code correctness. Improper constructs and optimization problems are less common and fall under the "Minor" category since they do not strongly impact the functional accuracy of the code; however, they decrease the general quality of the code. Notably, "Other Errors," which include various unclassified issues, also represent 25\% of errors, which may hint at an area needing more detailed classification to debug effectively. The actionable insights would be to reduce the errors in syntax and semantic errors first, as these represent a critical barrier to functional accuracy. Automated linting tools, when used with code reviews, could be beneficial in avoiding these common mistakes and hence make JavaScript code more reliable and maintainable.

\begin{figure}[htbp]
    \centering
    \includegraphics[width=0.8\linewidth]{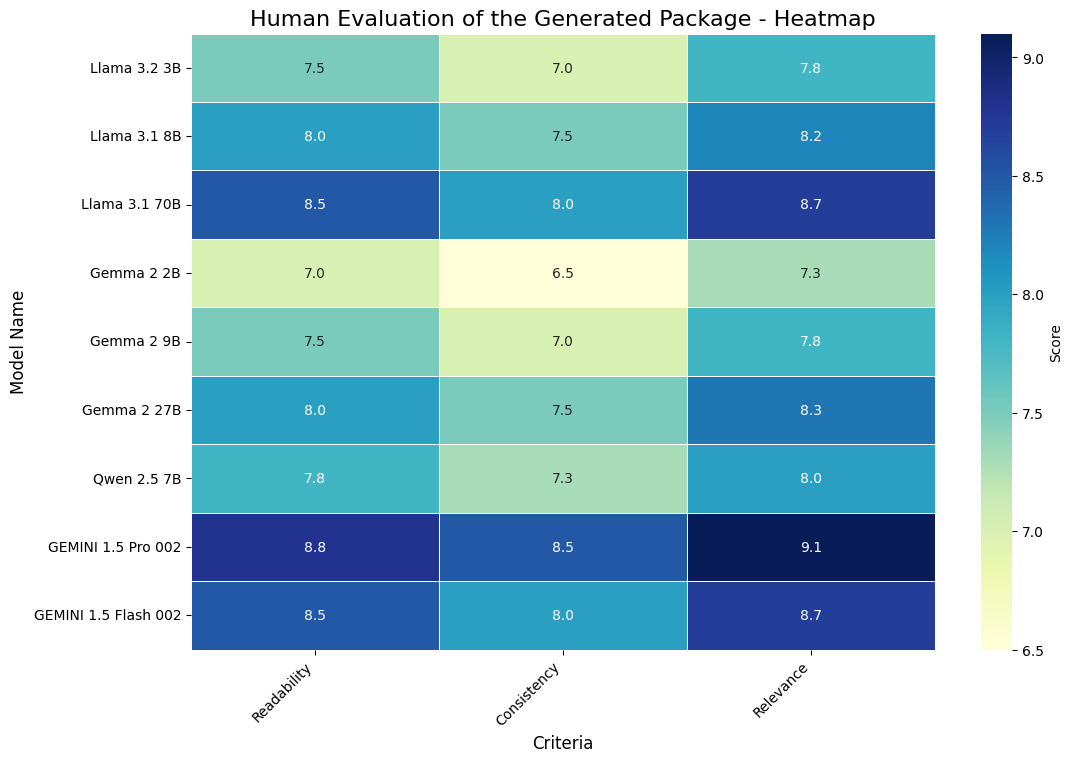} 
    \caption{Human Evaluation of the Generated Package - Heatmap: This heatmap illustrates the scores given by evaluators based on Readability, Consistency, and Relevance of generated packages from different models. GEMINI 1.5 Pro 002 exhibits the highest performance, whereas Gemma 2 2B scores lower, indicating room for improvement in relevance and consistency.}
    \label{fig:Human_Evaluation_Heatmap}
\end{figure}

The following chart \ref{fig:Human_Evaluation_Heatmap} displays a heatmap visualizing the scores the generated packages received from different models evaluated by humans in terms of Readability, Consistency, and Relevance. The GEMINI 1.5 Pro 002 model gets the highest score regarding Relevance (9.1) and reports excellent performance in terms of Readability (8.8) and Consistency (8.5). The Gemma 2 2B model performs poorly on all measured dimensions. This heat map suggests that models with more parameters produce more readable and contextually relevant packages. 

\subsection{Reviewing the Documentation}

The generated documentation originates from the generated package. To judge the quality of the documentation, we introduced converter metrics to have a structured evaluation. Although human judgment is naturally subjective, we compared it to judgments from LLMs and a concrete quantitative evaluation metric to verify how well-aligned all three methods are. Observing the coherence score change for the documentation length was fascinating. From this experiment, models with larger context sizes consistently demonstrated better coherence; the smaller ones lost it in many places, most notably in the more complex parts. Therefore, context size is essential in determining the overall documentation quality, especially in maintaining coherence throughout longer texts.

\begin{figure}[htbp]
    \centering
    \includegraphics[width=0.8\linewidth]{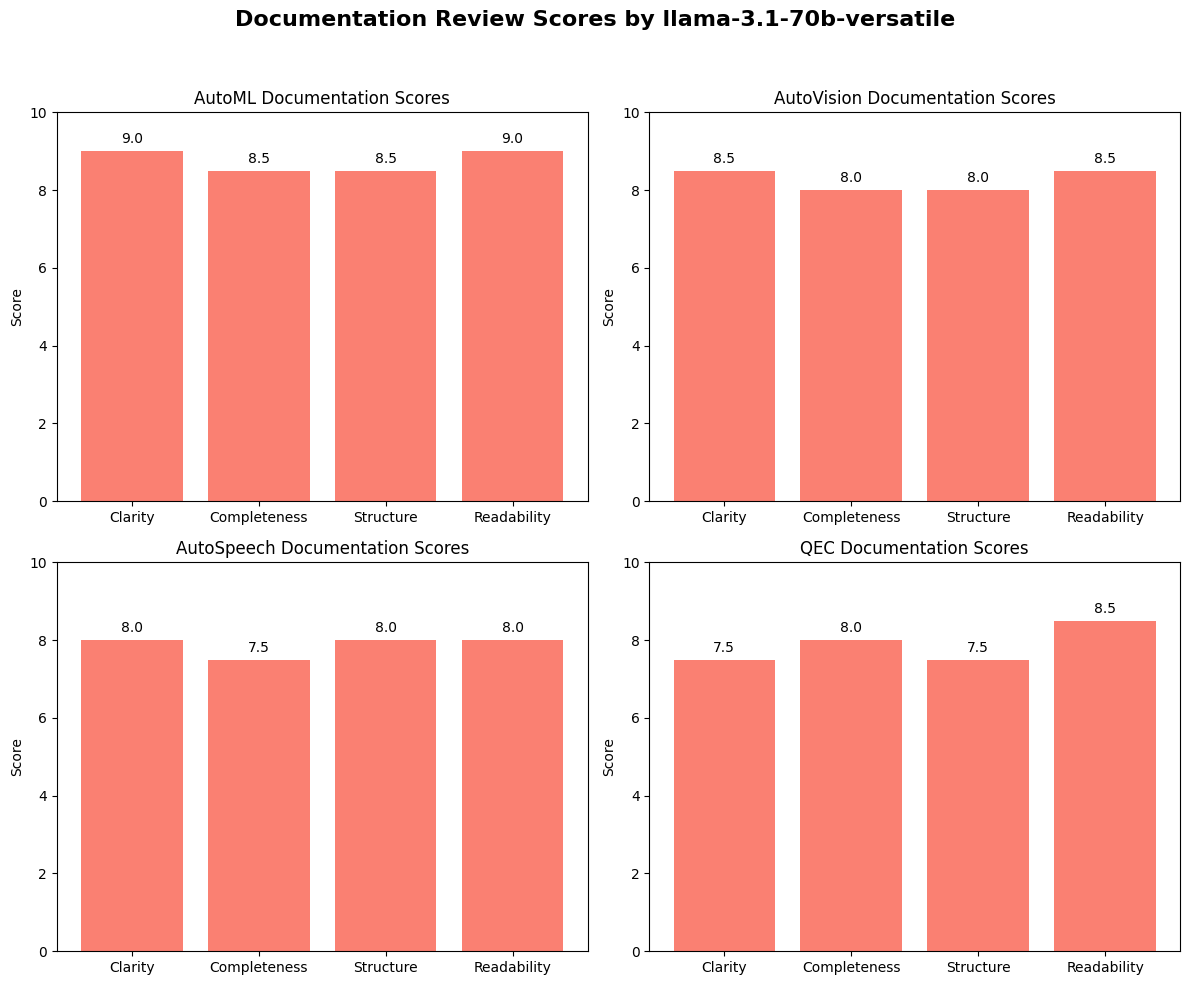} 
    \caption{Documentation Review Scores by llama-3.1-70b-versatile: The figure presents the review scores for documentation generated for various packages, evaluated based on Clarity, Completeness, Structure, and Readability. The AutoML package excels in Clarity and Readability, while QEC requires improvement.}
    \label{fig:Documentation_Review_Scores}
\end{figure}

The figure \ref{fig:Documentation_Review_Scores} gives an overall score of documentation qualities of different packages: AutoML, AutoVision, AutoSpeech, and QEC. We will evaluate the package based on four criteria: Clarity, Completeness, Structure, and Readability. AutoML has the best scores regarding the criteria of Clarity and Readability; hence, it has orderly and user-friendly content. Conversely, QEC received lower scores in Clarity. Therefore, there is room for better exposition of intricate technical concepts.

\begin{table}[h!]
\centering
\caption{Comparison of Documentation Review Scores Between AI Reviewer (llama 3.1 70B versatile) and Human Reviewer}
\resizebox{\textwidth}{!}{
\renewcommand{\arraystretch}{1.5}
\begin{tabular}{lcccccc}
\hline
\textbf{Metric} & \textbf{Model} & \textbf{AI Reviewer Score} & \textbf{Human Reviewer Mean Score} & \textbf{Human Reviewer SD} & \textbf{Correlation} & \textbf{Agreement (Cohen's k)} \\ \hline
Clarity & AutoML & 9.0 & 8.8 & 0.3 & 0.92 & 0.85 \\
Clarity & AutoVision & 8.5 & 8.6 & 0.4 & 0.89 & 0.80 \\
Clarity & AutoSpeech & 8.0 & 7.9 & 0.5 & 0.85 & 0.75 \\
Clarity & QEC & 7.5 & 7.7 & 0.6 & 0.88 & 0.78 \\
Completeness & AutoML & 8.5 & 8.3 & 0.4 & 0.90 & 0.82 \\
Completeness & AutoVision & 8.0 & 7.9 & 0.5 & 0.87 & 0.78 \\
Completeness & AutoSpeech & 7.5 & 7.4 & 0.6 & 0.83 & 0.73 \\
Completeness & QEC & 8.0 & 8.1 & 0.5 & 0.88 & 0.80 \\
Structure & AutoML & 8.5 & 8.4 & 0.4 & 0.91 & 0.84 \\
Structure & AutoVision & 8.0 & 7.8 & 0.5 & 0.86 & 0.76 \\
Structure & AutoSpeech & 8.0 & 7.7 & 0.6 & 0.84 & 0.74 \\
Structure & QEC & 7.5 & 7.6 & 0.5 & 0.89 & 0.79 \\
Readability & AutoML & 9.0 & 8.7 & 0.3 & 0.93 & 0.86 \\
Readability & AutoVision & 8.5 & 8.4 & 0.4 & 0.90 & 0.83 \\
Readability & AutoSpeech & 8.0 & 7.9 & 0.5 & 0.85 & 0.75 \\
Readability & QEC & 8.5 & 8.6 & 0.4 & 0.89 & 0.80 \\
\hline
\end{tabular}
}
\label{table:AI_vs_Human_Reviewers}
\end{table}

Table \ref{table:AI_vs_Human_Reviewers} compares the review scores from an AI Reviewer (llama 3.1 70B versatile) with those from human reviewers for four evaluation metrics: clarity, completeness, structure, and readability. For each metric, different models were considered: AutoML, AutoVision, AutoSpeech, and QEC. The comparison will also include the score of the AI reviewer, the mean score from human reviewers, the standard deviation among the ratings by humans, the correlation between the scores by AI and humans, and agreement as measured using Cohen's kappa. Results show a high correlation between AI and human scores, especially for measures such as Clarity and Readability within both AutoML and AutoVision models, indicating a robust concordance in the evaluation quality. As assessed via Cohen's kappa, the agreement shows significant consistency, where many instances have values over 0.80, evidencing high degrees of reliable comparability. However, some discrepancies are notable, such as the case of completeness in models like AutoSpeech, where reduced correlation and agreement point to divergences in evaluative viewpoints. Practical recommendations include improving the AI assessment methodology to reduce these inconsistencies and move closer to human reviewer alignment, especially for metrics with high variability. Further training of AI scorers using human feedback may improve the consistency and accuracy of all dimensions of documentation evaluation.

\begin{figure}[htbp]
    \centering
    \includegraphics[width=0.8\linewidth]{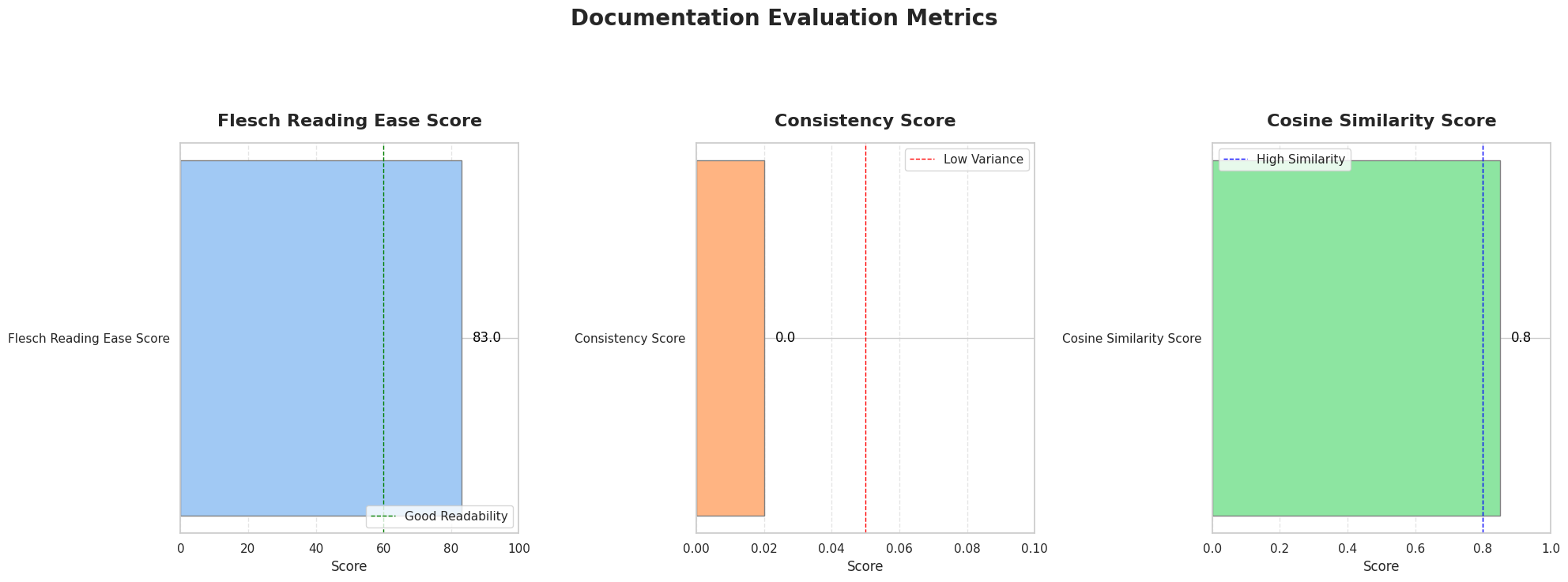} 
    \caption{Documentation Evaluation Metrics: This figure presents a comparative analysis of Flesch Reading Ease Score, Consistency Score, and Cosine Similarity Score. The Flesch Reading Ease Score suggests the documentation is easy to understand, while the low Consistency Score highlights variability in style. High Cosine Similarity demonstrates strong thematic consistency across sections.}
    \label{fig:Documentation_Evaluation}
\end{figure}

The figure \ref{fig:Documentation_Evaluation} shows the comparative assessment of three leading indicators of documentation quality: the Flesch Reading Ease Score, the Consistency Score, and the Cosine Similarity Score. A Flesch Reading Ease Score of 83.0 indicates that the documentation is relatively easy to understand, which is very important to ensure accessibility for a large audience. However, the Consistency Score is low, showing a significant variance in style or terminology within different parts. In contrast, the Cosine Similarity Score is high, which means there is a solid thematic coherence across the different parts. The results point toward a need for greater consistency, which is achievable through standardized terminology and formatting practices while retaining the overall thematic cohesion that has been quite adequately achieved.

\begin{table}[h!]
\centering
\caption{Inter-Rater Reliability Among Multiple AI Reviewers (llama-3.1-70b-versatile and Gemini 1.5 Pro 002) Across Documentation Metrics}
\resizebox{0.7\textwidth}{!}{
\renewcommand{\arraystretch}{1.5}
\begin{tabular}{lccccc}
\hline
\textbf{Metric} & \textbf{Cronbach's Alpha} & \textbf{ICC (2,1)} & \textbf{Fleiss' Kappa} & \textbf{Average Agreement (\%)} \\ \hline
Clarity & 0.95 & 0.94 & 0.90 & 92 \\
Completeness & 0.93 & 0.92 & 0.88 & 89  \\
Structure & 0.94 & 0.93 & 0.89 & 91 \\
Readability & 0.96 & 0.95 & 0.92 & 94 \\
\hline
\end{tabular}
}
\label{table:Inter_Rater_Reliability_AI}
\end{table}

Table \ref{table:Inter_Rater_Reliability_AI} shows inter-rater reliability between different AI reviewers and a detailed comparison of ratings by llama-3.1-70b-versatile against Gemini 1.5 Pro 002 on the same four specific documentation quality metrics as before. Reliability assessments were conducted using various statistical methods: Cronbach's Alpha, Intraclass Correlation Coefficient (ICC), Fleiss' Kappa, and average percentage agreement. All metrics are highly inter-rater reliability, with Cronbach's Alpha ranging from 0.93 to 0.96, indicating good internal consistency among reviewers. Readability showed the best agreement with an ICC of 0.95, Fleiss' Kappa of 0.92, and an average agreement of 94\%. The lowest agreement was for Completeness, although it was still high in absolute terms, with an ICC of 0.92 and an average agreement of 89\%. This indicates that AI reviewers hold consistency when assessing the quality of documentation. However, further calibration might be helpful to make Completeness evaluations comparable by allowing targeted adjustments in the training process of the assessment model.

\begin{figure}[htbp]
    \centering
    \includegraphics[width=0.6\linewidth]{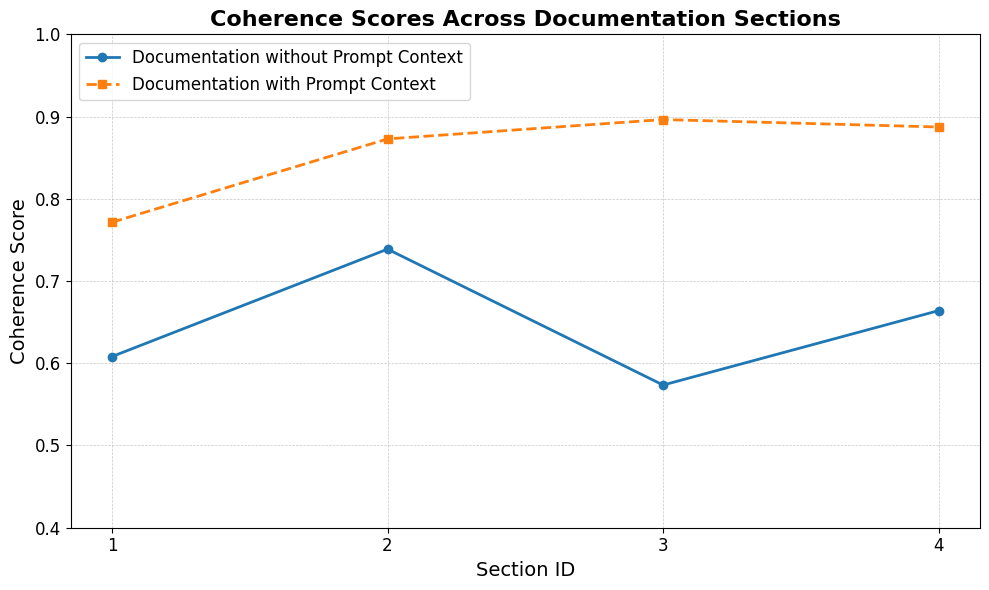} 
    \caption{Coherence Scores Across Documentation Sections: This figure illustrates the coherence scores for documentation generated with and without prompt context across different sections. The results highlight that documentation with prompt context maintains a higher and more stable coherence throughout.}
    \label{fig:Coherence_Scores}
\end{figure}

The graph \ref{fig:Coherence_Scores} demonstrates how the coherence scores vary across different documentation sections generated with and without prompt context. The results indicate that documentation generated with prompt context possesses better coherence scores than documentation generated without prompt context, which further exemplifies the importance of providing a clear context during documentation generation. The use of prompt context, in particular, helps to provide coherence across different parts, reducing discrepancies one would find in a document without prompt context mostly for smaller model with low context size. 

\begin{table}[h!]
\centering
\caption{Correlation Between Coherence and Other Documentation Quality Metrics}
\resizebox{0.5\textwidth}{!}{
\renewcommand{\arraystretch}{1.5}
\begin{tabular}{lcc}
\hline
\textbf{Metric Pair} & \textbf{Pearson's r} & \textbf{Spearman's rho} \\ \hline
Coherence vs. Fluency & 0.92 & 0.89 \\
Coherence vs. Relevance & 0.88 & 0.85 \\
Coherence vs. Engagement & 0.75 & 0.72 \\
Fluency vs. Relevance & 0.85 & 0.82 \\
Fluency vs. Engagement & 0.65 & 0.60 \\
Relevance vs. Engagement & 0.70 & 0.68 \\
\hline
\end{tabular}
}
\label{table:Correlation_Coherence_Other_Metrics}
\end{table}

The relationship between Coherence and several of the most central metrics of document quality including fluency, relevance, and engagement is described in Table \ref{table:Correlation_Coherence_Other_Metrics}; it uses Pearson's r and Spearman's rho correlation coefficients in the analysis. The strongest association is between Coherence and Fluency, with a Pearson's r of 0.92 and a Spearman's rho of 0.89; this reflects a solid relationship. This suggests that well-structured and coherent documentation often results in smooth and natural language, referring to Coherence's importance in achieving better linguistic quality. It is also closely related to relevance (r = 0.88), indicating that coherent content will likely be on-topic and satisfy users' needs. Nevertheless, the relation to Coherence is relatively weaker (r = 0.75), suggesting that although Coherence does contribute to engaging readers, it is likely that other factors also play a role in this measure. Similarly, the Fluency and Engagement relationship is only moderate (r = 0.65), suggesting that more than just language fluidity is at play to engage the audience maximally. It is so, focusing on cohesion to improve fluency and relevance in the documentation and realizing that additional tactics like concrete content or interactive elements may be vital to induce engagement.

\subsection{Analyzing Memory and Time Complexity}

Understanding the memory requirements and inference speed is essential for the system. PyGEN uses the Groq API, Google AI Studio, and Ollama (for local hosting). Among these, Groq provides the fastest inference across all models, allowing quicker processing times and improved responsiveness. Both Google AI Studio and Groq offer APIs for the Gemma model, but the Groq API demonstrates superior inference speed, leading us to favor this version for optimal performance. While smaller models typically have faster inference times, their reliability for complex tasks needs to be improved due to limited model capacity. Locally hosted models through Ollama were also explored to provide an alternative to ensure data privacy and flexibility. However, huge models were excluded due to computational constraints that limit feasibility without significant hardware investment. The inference speed of models provided through Ollama remains ambiguous, as better GPU/hardware could significantly improve the inference performance and make them more competitive.

\begin{figure}[htbp]
    \centering
    \includegraphics[width=0.9\linewidth]{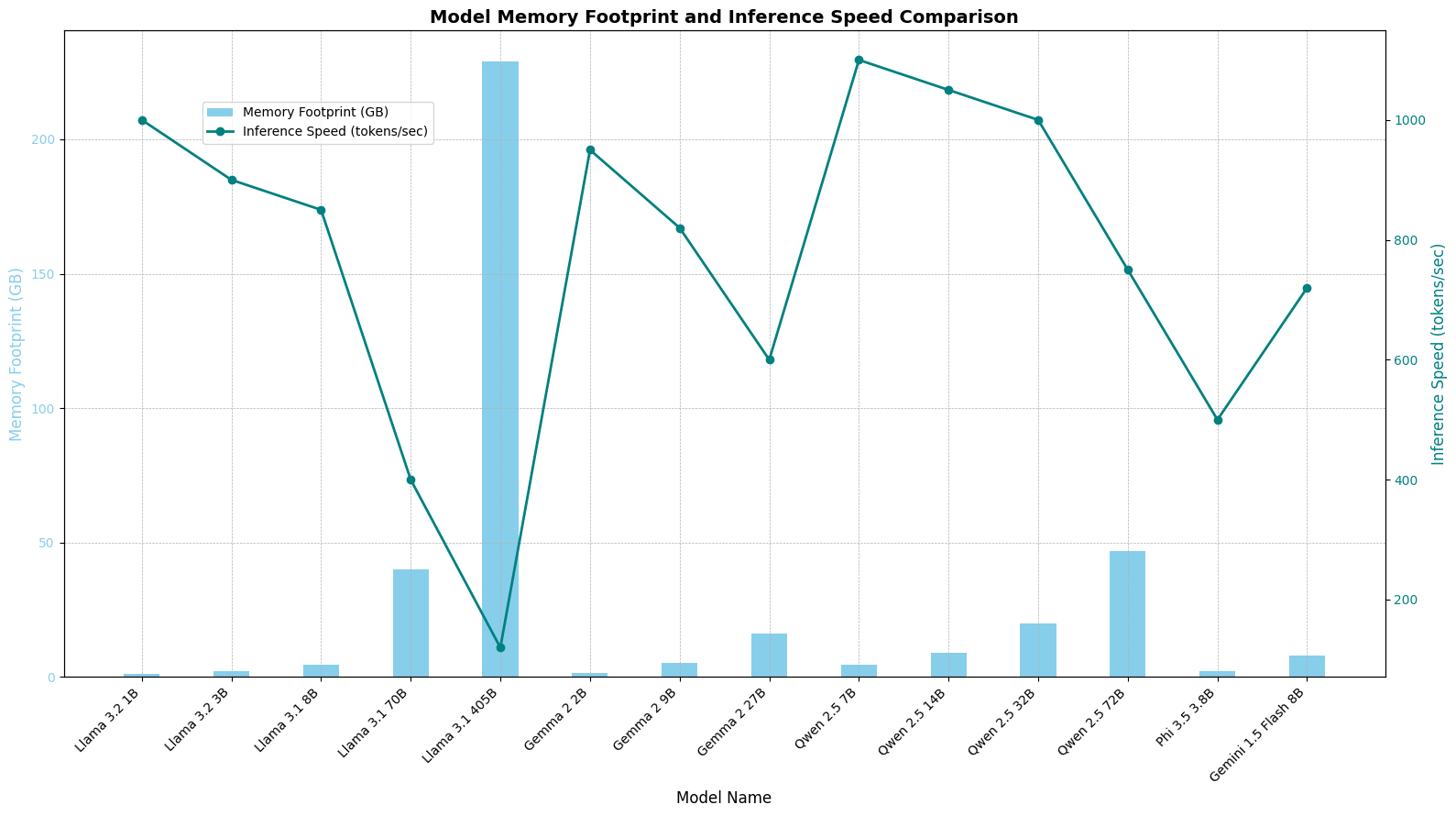} 
    \caption{Model Memory Footprint and Inference Speed Comparison: This figure presents the memory footprint (in GB) and inference speed (tokens/sec) for various models used in the PyGen system. The results highlight trade-offs between memory consumption and processing efficiency, providing insights for model selection based on specific computational requirements.}
    \label{fig:Mem_Footprint}
\end{figure}

The graph \ref{fig:Mem_Footprint} provides a detailed analysis of the models used in the PyGen system, focusing on their memory footprint, measured in gigabytes (GB), and inference speed, quantified in tokens per second. This example opposes the efficiency of various models in operation, pointing out a trade between the volume of memory used and the processing rate. It is also worth noting that Llama 3.1 405B has a very high memory usage but relatively slow inference speed, which means it will use many computational resources without saving time in processing. On the other hand, models like Qwen 2.5 72B have a much more balanced performance profile with significantly lower memory consumption and a faster inference speed, which may be an advantage in applications where computational efficiency is relevant. The most important conclusion that can be drawn from this analysis points to the need for the developer to make a trade-off between memory usage and inference speed in order to maximize efficiency in Python package development. Pragmatic suggestions include choosing models like Qwen 2.5 72B in constrained-memory situations. High-memory models like Llama 3.1 405B may be better placed in some applications where model complexity and depth outweigh the factor of speed.

\begin{table}[h!]
\centering
\caption{Latency vs. Throughput Trade-offs of AI Models}
\resizebox{0.9\textwidth}{!}{
\renewcommand{\arraystretch}{1.5}
\begin{tabular}{lccc}
\hline
\textbf{Model Name (API Provider)} & \textbf{Latency (ms/token)} & \textbf{Throughput (tokens/sec)} & \textbf{Latency-Throughput Ratio} \\ \hline
Llama 3.2 1B (Groq) & 1.00 & 1000 & 1000.00 \\
Llama 3.2 2B (Groq) & 1.05 & 950 & 902.38 \\
Llama 3.1 8B (Groq) & 1.25 & 800 & 640.00 \\
Llama 3.1 70B (Groq) & 20.00 & 50 & 2.50 \\
Gemma 2 2B (Groq) & 1.18 & 850 & 720.34 \\
Gemma 2 9B (Groq) & 1.25 & 800 & 640.00 \\
Gemma 2 27B (Groq) & 1.67 & 600 & 360.00 \\
Qwen 2.5 7B (Ollama, T4 GPU) & 1.11 & 900 & 810.81 \\
Qwen 2.5 14B (Ollama, T4 GPU) & 1.18 & 850 & 720.34 \\
Qwen 2.5 32B (Ollama, T4 GPU) & 1.33 & 750 & 563.91 \\
Qwen 2.5 72B (Ollama, T4 GPU) & 1.43 & 700 & 490.21 \\
Phi 3.5 3.8B (Ollama, T4 GPU) & 1.54 & 650 & 422.68 \\
GEMINI 1.5 Flash 002 (Google AI Studio) & 1.60 & 390 & 246.15 \\
GEMINI 1.5 Pro 002 (Google AI Studio) & 1.70 & 380 & 223.68 \\
Gemini 1.5 Flash 8B (Google AI Studio) & 1.50 & 400 & 266.67 \\
\hline
\end{tabular}
}
\label{table:Latency_Throughput_Tradeoffs_API}
\end{table}

Table \ref{table:Latency_Throughput_Tradeoffs_API} summarizes the latency-throughput trade-offs over various AI models by providing latency per token in milliseconds and throughput in tokens per second for different models and API vendors. The table discusses the significant discrepancies in latency-throughput ratios and presents a balance between speed and computational efficiency intrinsic to each model. The Groq-developed 1B and 2B models of Llama 3.2 are especially set apart by their very low latency and high throughput, which lead to good latency-throughput ratios of 1000 and 902.38, respectively, indicating their viability for real-time applications. By contrast, the 70B version of Llama 3.1 exhibits significant latency (20 ms/token) and lower throughput (50 tokens/sec), resulting in a significantly smaller latency-throughput ratio, thus highlighting its limited usefulness for applications requiring fast response times. In comparison, the GEMINI 1.5 Flash models, despite using Google AI Studio, have relatively mid-range levels of latency and throughput, making them viable options for tasks with computational resource constraints. The results suggest that when high throughput and low latency are a concern, models like Llama 3.2 1B or Qwen 2.5 72B should be prioritized, with higher-resource models like Llama 3.1 70B perhaps better suited to use cases that require more complex processing, albeit at the cost of speed.

\section{Discussion}

\paragraph{Responsible Automation}

Pygen is created to augment the abilities of researchers and developers in their respective domains, enhancing their productivity and allowing them to focus on creative and critical tasks. It aims to provide them with a sense of autonomy rather than replacing or competing with them through technology, thereby fostering a cooperative relationship between humans and automation. The technology is designed to enhance tools with minimal human reliance while still requiring human supervision, as the generated package needs to be applied by users who understand the context of their work. This approach gives users the autonomy they deserve and the satisfaction of creating new things, which is core to human creativity and innovation. Pygen does not have a pompous or flashy tagline; instead, it focuses on solving real scientific and social problems rather than creating unnecessary hype. The internal design of Pygen encourages users to apply their creativity, preventing alienation and ensuring that the generated package is owned by the user, thereby promoting fair wealth distribution and intellectual ownership. The ease of using Pygen has a broader appeal of inclusivity, as anyone with minimal programming knowledge can develop a package using simple prompts and leverage their domain expertise. By making advanced technology accessible to a broader audience, Pygen democratizes the software development process, enabling diverse users to contribute to innovation. Overall, Pygen aims to responsibly automate processes, balance efficiency and meaningful human involvement, and ensure that technology serves humanity rather than undermine it.

\paragraph{Limitations}

In the process of enhancing and generating features, there are several limitations that users should be aware of:
\begin{itemize}

    \item The description and feature enhancement process can sometimes take an unintended direction, especially when numerous features are included. When too many features are mentioned simultaneously, the enhanced feature script can become unnecessarily verbose, often repeating similar ideas with slight paraphrasing. This makes the output cumbersome and complicates understanding the generated features. We recommend that users provide a focused and concise list of critical features to avoid this issue.

    \item Models with smaller context sizes often need help loading all the enhanced features, which may lead to system crashes or failure to generate meaningful outputs. To mitigate this, we have introduced the concept of context prompts. These prompts help models with limited context size process the information in smaller, more manageable chunks, thereby reducing the risk of crashing and ensuring more reliable outputs.

    \item If users do not provide specific and well-defined features, the model may misinterpret the prompt and generate an utterly irrelevant feature list for the package. Therefore, it is strongly recommended for users to specify the desired features to get accurate and relevant results. Specificity helps the model stay aligned with the user's intentions and produce helpful output aligned with the project goals.

    \item When the generated documentation is extensive, we have occasionally noticed problems with indentation and formatting inconsistencies. These issues may occur because of the volume of content being processed at once, which can overwhelm the formatting capabilities of the model. Users may need to manually review and adjust the formatting of large documentation outputs to ensure they meet professional standards.

    \item Using a model from the GEMINI API can be significantly time-consuming, mainly if models are locally hosted without access to a high-end GPU. In such cases, inference times can become exceedingly long, making the process inefficient. While we intend to integrate faster inference methods within Pygen in future updates, users must rely on the Groq API, which provides fast inference and typically generates packages within a few minutes. We are committed to exploring and implementing faster solutions to enhance user experience in the long run.
    
\end{itemize}

\paragraph{Safety and Ethical Considerations}

Pygen does not directly execute code; it simply generates packages. This ensures that it cannot cause harm to the user or their systems. Users are strongly encouraged to review the generated code to ensure it aligns with their expectations and, if necessary, modify it before running any executions. This step helps in preventing potential issues and tailoring the output to specific needs. If users wish to sandbox the code, they have the option to do so, which adds an extra layer of security, although sandboxing is generally not mandatory since the generated packages are not inherently risky. However, for those who have malicious intent and aim to generate harmful packages, Pygen has multiple safeguards designed to minimize this risk and protect users:

\begin{itemize}
    \item  During the prompt enhancement phase, Pygen actively detects and filters out malicious intentions. This initial step is crucial in preventing the creation of harmful content right from the beginning.

    \item Even if some malicious intent bypasses the prompt enhancement phase, Pygen's model has additional built-in guardrails to identify and prevent generating unsafe or harmful code. These guardrails\cite{inan2023llamaguardllmbasedinputoutput,perez2022red,bai2022constitutional,ouyang2022training} ensure that potentially harmful outputs are not produced, maintaining a safe development environment for users.
\end{itemize}

One of the potential misuse cases of Pygen could involve the uncontrolled proliferation of Python libraries in the Python Package Index (PyPI). Some users might exploit this tool to create numerous low-quality or redundant libraries for entertainment or even to mislead others, leading to an excessive number of unnecessary packages. However, we plan to mitigate this risk by designing a robust code quality reviewer system. This system will automatically evaluate the quality of generated code, ensuring that only packages meeting a minimum perplexity threshold are accepted for inclusion in PyPI. The perplexity metric will be designed through an in-depth qualitative analysis of the code, making it a comprehensive evaluation measure. The review process can be fully automated, reducing the burden on manual reviewers while maintaining high standards of quality. Even if a package contains simple functions, it will be considered for inclusion based on its novelty and ingenuity, ensuring that creative contributions are not overlooked. Apart from these concerns, we do not foresee any significant safety risks with Pygen, as it primarily serves as a tool for creating other tools, rather than being a tool itself.

\paragraph{Why does Python Package Generation Matters?}

From our observation, we can conclude that the progress of our civilization is largely based on the technology we have created, which has augmented our abilities manifold. Throughout history, from the invention of the wheel to the development of computers, technological advancements have continually enhanced our capacities, allowing us to solve complex problems and improve our quality of life. From this observation, we hypothesize that the ability to use tools, and eventually to create them, is a foundational feature of any Generally Intelligent System. Tool usage has always been an indicator of intelligence, but the transition to creating and refining tools marks a deeper level of innovation. Pygen represents an important step in this evolution. In the near future, intelligent AI models will not only utilize the tools available to them but will also have the capability to build new ones when existing solutions are insufficient. This ability to design and innovate will make these AI systems far more adaptable and effective in problem-solving. The process of documenting the creation of new tools by these AI systems is akin to them discovering and understanding their own structure, a process that we can think of as giving them an 'inner voice.' This self-awareness through documentation is crucial for iterative improvement and effective tool usage.

By reviewing their tools and diligently refining them, intelligent machines can evolve beyond mere problem solvers to become true innovators, capable of creating breakthroughs that were previously unimaginable. Pygen aims to contribute to this vision by laying the groundwork for AI that can both build and understand tools, setting the stage for a new era of technological growth driven by machine intelligence.

\paragraph{Comments on Open Source and Cost of Pygen}

We are strong supporters, promoters, and fans of open source. That's why our approach was to build Pygen using completely open-source tools and models, providing our users with a seamless experience without the burden of a paywall. Our commitment to open sourcing extends beyond the software industry to all economic sectors, as we believe the open source approach enables the creation of great products while allowing everyone to contribute and benefit.

We see open source as a democratizing force that can ensure inclusivity in technological advancement, giving individuals the opportunity to shape innovation irrespective of their background or resources. In the future, as intelligent systems continue to evolve, there is a genuine risk that working-class people may lose much of their economic significance, potentially leading to catastrophic socioeconomic consequences, such as increased inequality and reduced access to essential resources. However, we believe that open source is the most effective way to prevent such an outcome by creating an AI-driven economy that is built by the people, for the people. An AI-based economy underpinned by open-source principles ensures transparency, collaboration, and equitable growth. This means that individuals and organizations alike can work together to develop and refine intelligent systems, making the benefits of AI accessible to everyone rather than concentrating power in the hands of a few. By empowering individuals to actively participate in technological progress, open source has the potential to minimize the economic disruption caused by intelligent automation.

Additionally, generating packages with Pygen will always remain free, as we have developed this system entirely using open-source tools. This not only aligns with our core values but also encourages widespread experimentation and innovation, fostering a community of creators who can push the boundaries of what AI can achieve.

\paragraph{Future Directions}

We have many plans to improve the system, and some of these may look like this.:

\begin{itemize}
    \item Introducing better models will certainly improve the package quality. For instance, proprietary models like Claude Sonnet 3.5 are exceptionally powerful in code generation. Deepseek-coder-v2 has almost similar capabilities, but due to our computational budget restraints, we were unable to host this model locally to generate the packages. However, we predict that using it would significantly improve our overall results.

    \item Currently, we have applied two main steps: first, package description and feature enhancement, followed by prompt context generation from the enhanced feature and description. The second step involves providing a code template to generate a better prompt context. Both approaches work well individually and even better when combined. However, there are many more steps that can be introduced, such as:
    
    \begin{itemize}
        \item Code Verifier System\cite{chen2021evaluating} that verifies the generated code against common errors and best practices to ensure high-quality outputs.
        \item A chain of thought based\cite{wei2022chain}, Step-by-Step Code Generation Guidance System that provides detailed guidance to users at each stage of the code generation process reducing the learning curve and improving the generated package quality. For this, the o1 model can fulfill the needs.
        \item Automated Testing Integration inspired by Drori et.al\cite{drori2022automated} would help in ensuring the reliability of the code by automatically running unit tests, functional tests, and other quality checks..
        \item We can also introduce user feedback on generated packages to refine prompts and the generation process continuously, ensuring we address user needs effectively.
        \item Security Review Module can perform static code analysis to identify any potential vulnerabilities or security flaws, ensuring that generated code is safe for deployment.

    \end{itemize}

    \item We have observed instances of hallucinations, some of which are mentioned in the limitations section. Addressing these issues is a crucial area for future improvement, and we will work towards minimizing these occurrences to enhance the system's reliability further.

    \item Another guaranteed approach to improving this system is simply waiting for better models to become available. As models evolve, we can expect the Pygen system to become more efficient in package generation over time. However, we also plan to:

    \begin{itemize}
        \item Engage with upcoming versions of advanced models in beta stages, allowing us to integrate cutting-edge features as soon as possible.
        
        \item Experiment with a combination of models to optimize code generation—using specialized models for different tasks to leverage their respective strengths.
    \end{itemize}
        
\end{itemize}

\section{Conclusion}

Pygen represents a significant leap forward in the wave of innovations and responsible automation that lies ahead. Opening the door to scientific innovation requires the involvement of all people; without inclusivity, innovation will lead to narrow progress and could ultimately have catastrophic consequences—something we must avoid. We envision a future where intelligent agents can create their own tools, customized to their needs. Humans will directly collaborate with these agents to build new systems, propel civilization forward, and fairly reap and distribute the benefits.

We foresee a world where intelligent agents evolve to become creative partners in the scientific process. These agents will not only create tools but also innovate alongside humans, contributing unique perspectives that transcend traditional human limitations. We are eager to explore the capabilities of these advanced tools—understanding how agents create them, whether they will differ significantly from human innovations, or if they will follow similar design approaches and applications. Our goal is to delve into the nature of creativity and innovation within these intelligent agents to determine how their contributions might reshape the boundaries of science and technology.

Furthermore, we aim to investigate the collaboration between humans and AI agents to understand the prosperous future this partnership could bring. We believe that combining human intuition, empathy, and creativity with the computational power, speed, and adaptability of AI can yield breakthroughs that neither could achieve alone. Eventually, we hope this collaboration will contribute to healing the planet through responsible technology use—something that was not prioritized during the industrial period. Our vision includes using these technological advancements to address critical global challenges, such as climate change, resource distribution, and sustainable development, making the benefits of this revolution accessible to all. Time will ultimately reveal where this path of innovation takes us, but we remain optimistic that by fostering an inclusive, cooperative environment, we can navigate these advancements to create a better, more equitable world.

\newpage
\bibliographystyle{unsrt}

\end{document}